\documentclass[12pt,noshowpacs,nofootinbib,notitlepage,amsmath]{revtex4-1}
\usepackage{setspace}
\allowdisplaybreaks
\usepackage{graphicx,color}
\usepackage[colorlinks=true,citecolor=blue,linkcolor=blue,urlcolor=blue]{hyperref}
\usepackage[charter]{mathdesign}
\usepackage{multirow}
\usepackage{enumerate}

\DeclareSymbolFontAlphabet{\mathcal}{symbols}
\DeclareSymbolFont{symbols}{OMS}{xmdcmsy}{m}{n}
\DeclareSymbolFont{largesymbols}{OMX}{xmdcmex}{m}{n}
\SetSymbolFont{symbols}{bold}{OMS}{xmdcmsy}{b}{n}

\begin{document}  
\title{\color{blue}\Large Not quite black holes at LIGO}

\author{Bob Holdom}
\email{bob.holdom@utoronto.ca}
\affiliation{Department of Physics, University of Toronto, Toronto, Ontario, Canada  M5S 1A7}
\begin{abstract}
We provide more evidence of not quite black holes at LIGO. We update and streamline our previous search strategy and apply it to the ten black hole merger events and the one neutron star merger event. The strategy is aimed at the evenly spaced resonance spectrum expected from not quite black holes, given that at low frequencies the radial wave equation describes the modes of a stretched 1D cavity. We describe various indications of the self-consistency of the apparent signals across all events in the context of a simple theoretical model. The merger with the largest final mass, spin and redshift, GW170729, provides additional interesting support.
\end{abstract}
\maketitle

\section{Introduction}

A ``not quite black hole'' is a horizonless object where strong gravity extends out to a radius just slightly beyond, by a distance $\delta r$, the would-be horizon. Such an object is very close to being a black hole since the distance $\delta r$ is characterized by the Planck length. The possibility of having something other than a black hole (BH) as the end point of gravitational collapse is related to having a UV complete theory, since the interior of these objects probe the UV completion. If this UV completion involves an action with terms quadratic in the curvature then these not quite BH solutions exist and are referred to as ``2-2-holes'' \cite{Holdom:2002xy,Holdom:2016nek}. At the same time quadratic gravity is an old candidate for a UV complete quantum field theory of gravity, since as such it shares with QCD the properties of renormalizability and asymptotic freedom \cite{stelle}. Other properties of the quantum theory are still being elucidated.

Recently we found 2-2-hole solutions that are sourced by an ordinary relativistic gas \cite{Holdom:2019ouz} (for further analysis see \cite{Ren:2019jft}). The total entropy of the gas turns out to satisfy an area law, and this entropy is somewhat larger than the entropy of a black hole of the same mass. In this way 2-2-holes may be preferred as the end point of gravitational collapse. Like BHs, they have no upper limit on their size. Unlike BHs, they have a minimum size where they are cold, and these Planck-size objects have an internal structure quite different than the large variety. This situation is not unlike QCD. The two types of solutions are analogous to the hadrons and the quark matter states of QCD. The latter states show that the effects of strong interactions can extend over macroscopically large regions even though the fundamental scale of strong interactions is microscopic. In the classically scale invariant version of quadratic gravity, the Planck scale can arise as the scale of strong interactions, just as the QCD scale arises in QCD \cite{Holdom:2015kbf}.

We thus consider the possibility that all BHs in nature are in fact not quite BHs. This is a conjecture that can be tested, since a gravitational wave signal can distinguish not quite BHs from BHs. The effective radial description of low frequency waves around the not quite BHs is that of a 1D cavity. One end of the cavity is at the origin ($r=0$) and the other end is at the angular momentum potential barrier ($r=3M$, in units with $G=1$). Since the potential barrier slightly leaks, a pulse that moves back and forth in the cavity can produce periodic pulses (echoes) observed on the outside \cite{Cardoso:2016rao,Cardoso:2016oxy}. This observation motivated searches for echoes in LIGO data, beginning with \cite{Abedi:2016hgu}. But the perturbed state of a not quite BH, when newly formed after a merger, may be more complicated. If it produces something other than a single pulse moving back and forth, then the simple echo waveform will be replaced by something else.

A 1D cavity has a more general feature: an evenly spaced resonance spectrum. This is a more robust and general search target than actual echoes \cite{Conklin:2017lwb,Conklin:2019fcs}. We can express the size of the cavity as a tortoise coordinate distance $\Delta x$. Then the spacing between resonances is $\Delta f=1/(2\Delta x)=1/\Delta t$ where $\Delta t$ is the round-trip travel time (the time delay between the possible echoes). The cavity size is stretched compared to the physical size of the not quite BH since the dimensionless ratio $\Delta t/M$ is large, something like 800 or so. Since an explicit solution for a rotating not quite BH does not exist yet, we instead  use a truncated Kerr spacetime as an approximation. The truncation means that a boundary or wall is introduced slightly outside the horizon. $\Delta x$ is now the distance between the wall and the potential barrier. This construction yields no angular momentum barrier at the wall, just as a 2-2-hole has no angular momentum barrier at its origin. We discuss the known formula for $\Delta t/M$ in this model in Section \ref{s3}.

We focus on the dominant $\ell=m=2$ gravitational wave mode and obtain the corresponding solution of the Sasaki-Nakamura (SN) equation, and thus obtain the function $\psi_\omega$ that encodes the spectrum at asymptotic infinity \cite{Conklin:2019fcs}. This is a product of a transfer function $K(\omega)$ and a source integral $D(\omega)$. The transfer function contains the resonance spectrum that depends on the mass $M$ and dimensionless spin $\chi$, along with $\Delta x$ and a quantity $R_{\rm wall}$ that describes the boundary condition. $R_{\rm wall}=-1$ is a purely reflecting boundary condition (and is the analog of the Dirichlet boundary condition found for the spinless 2-2-hole), but we choose a slightly less negative value, $R_{\rm wall}=-1+\epsilon$, to account for a small amount of damping or dissipation of the wave as it traverses the material inside the not quite BH.

\begin{figure}[t]
\centering
   \includegraphics[width=0.48\textwidth]{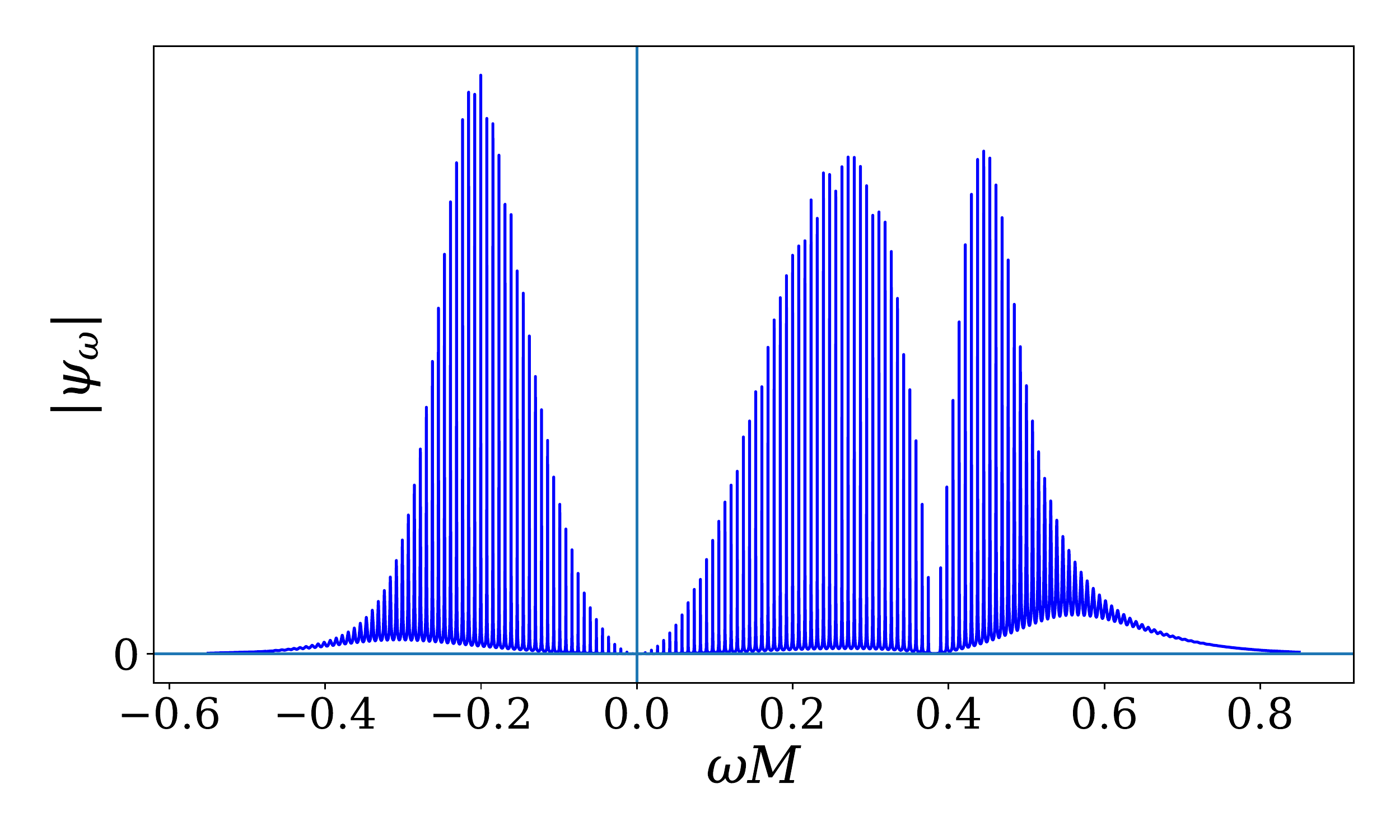} 
   \includegraphics[width=0.48\textwidth]{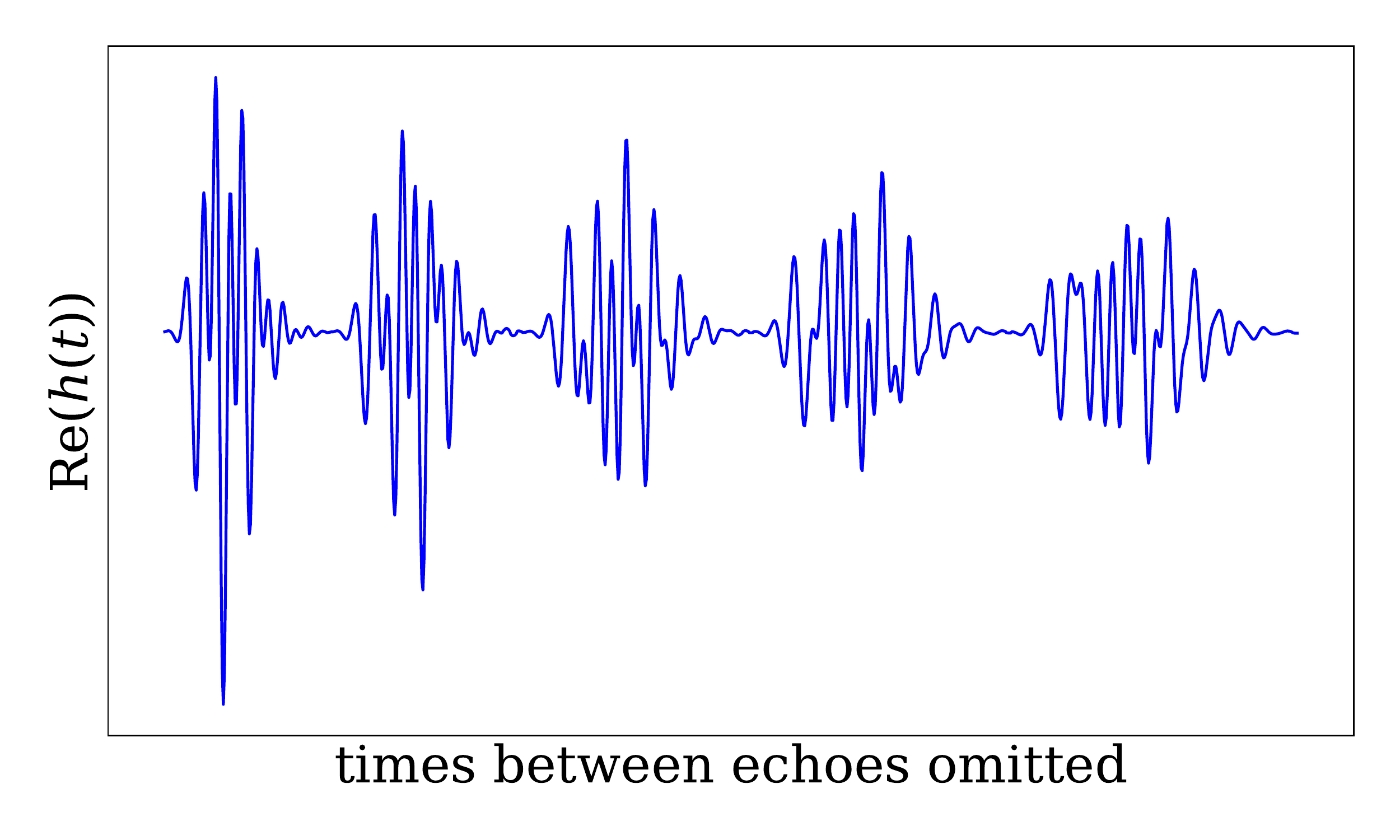}  
\caption{a) $|\psi_\omega|$ as obtained from the SN equation.  b) The first five echoes.}
\label{fig1}
\end{figure}

The source integral $D(\omega)$ modulates the resonance spectrum and its dependence on the initial condition was explored in \cite{Conklin:2019fcs}. The simplest initial condition is a single traveling pulse originating inside the cavity. The frequency content of the pulse can be characterized by a function $f(\omega)$ that in turn determines the spectral flux density
\begin{align}
\frac{dE}{d\omega}=8\omega(\omega-\omega_0)|f(\omega)|^2.
\end{align}
This is negative for $0<\omega<\omega_0$ and we choose $f(\omega)\propto \exp(-\kappa M^2\omega(\omega-\omega_0))$ according to the expectation that the lowest energy modes are the most likely to be excited. (With $m=2$, $\omega_0=\chi/r_+$, where $r_+$ is the horizon radius.) An example of $|\psi_\omega|$ is given in Fig.~\ref{fig1}a for the values $\chi=2/3$, $\Delta t/M=800$ and $\kappa=13$. Less trivial initial conditions can be modeled with more traveling pulses and will result in a less smooth envelope for the spikes. This envelope always vanishes at $\omega_0$.

From $\psi_\omega$ the strain waveform in the time domain $h(t)$ can be obtained by an inverse Laplace transform. As shown in \cite{Conklin:2019fcs}, $\psi_\omega$ contains a factor of a function $c_0(\omega)$ that appears in the transformation of Teukolsky to SN amplitudes. But to obtain the strain $h(t)$ from the output of the SN equation $\psi_\omega$, this output must first be divided by $c_0(\omega)$ before the inverse Laplace transform is performed \cite{Conklin:2019smy}, and thus the $c_0(\omega)$ factor is canceled. The resulting Re$(h(t))$ is shown in Fig.~\ref{fig1}b where we see irregular echoes for even the simplest of initial conditions.

In \cite{Conklin:2017lwb} a search strategy was developed to target the resonance spectrum. Some evidence for such spectra was presented for four BH merger events reported early by LIGO. Here we update this search and apply it to all ten confident detections of BH mergers in the first and second observing runs \cite{LIGOScientific:2018mvr}, using publicly available data \cite{data}. Our goal is to simplify the strategy and to apply it uniformly to all events. This reduces look-elsewhere-effects and it makes our analysis and results more easily reproducible. Another feature of our analysis is that an optimized bandpass, a purely data-driven quantity, can be compared to the predicted spectrum for each event.

\section{The search}

The search strategy involves taking a range of whitened data starting at the merger time and having duration $T$. Then after a FFT, the absolute value is taken. A complication is that the gravitational wave arriving at Earth has two polarizations which form the real and imaginary parts of a complex waveform. A LIGO detector, depending on its orientation, projects this complex waveform into a set of real numbers.

Thus to find the observable version of the signal spectrum, we first must model this projection, and we do this by taking the real part of the time-domain signal strain waveform $h(t)$. (Different projections give qualitatively similar results.)  Carrying out the steps of the search strategy on Re$(h(t))$ gives a ``reconstructed'' spectrum that we will label as $h(f)$. This quantity has an additional dependence on $T=N_E\Delta t$. $N_E$ is the number of echoes, loosely speaking, since typically any echoes will have already merged by the time $T$ is reached.

\begin{figure}[t]
\centering
   \includegraphics[width=0.48\textwidth]{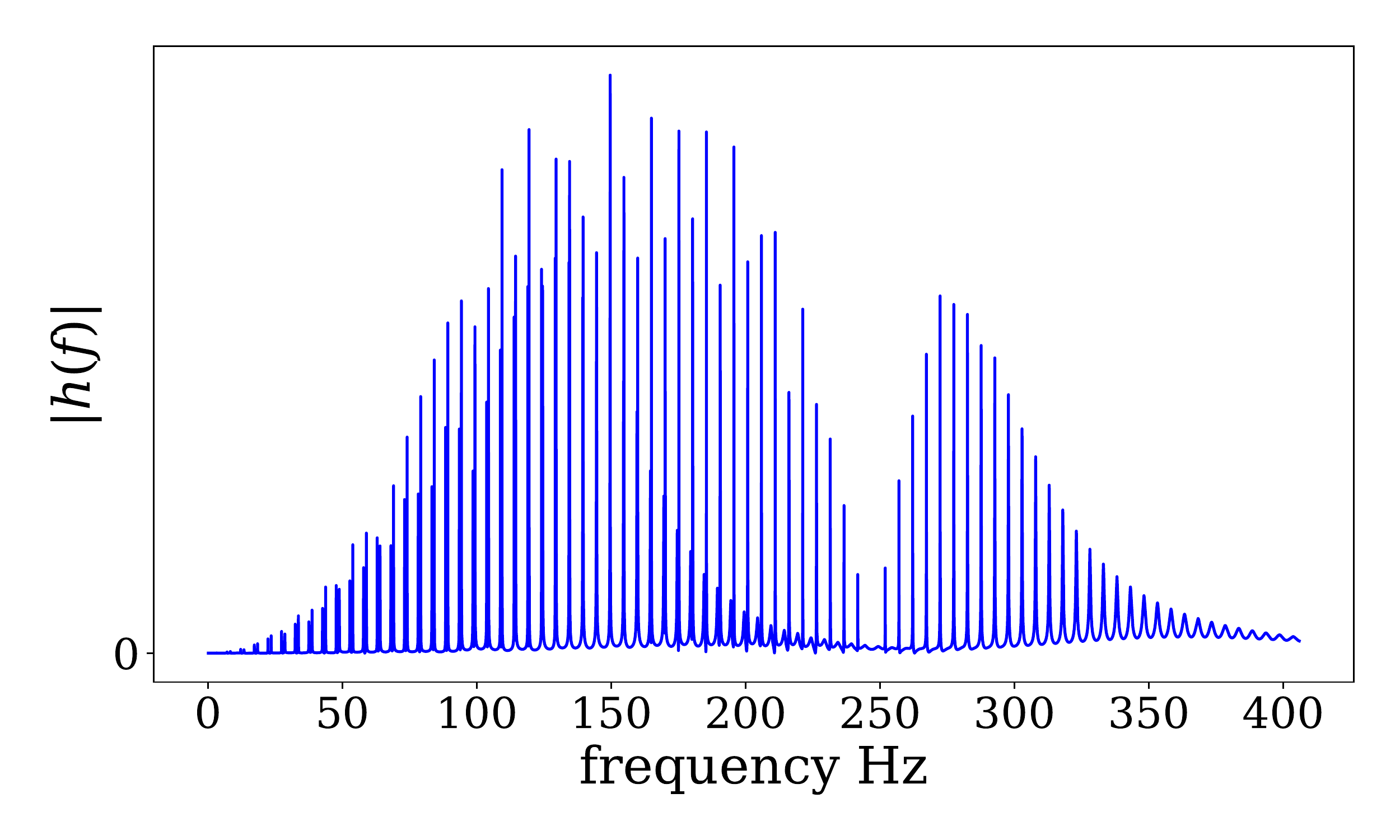}  
   \includegraphics[width=0.48\textwidth]{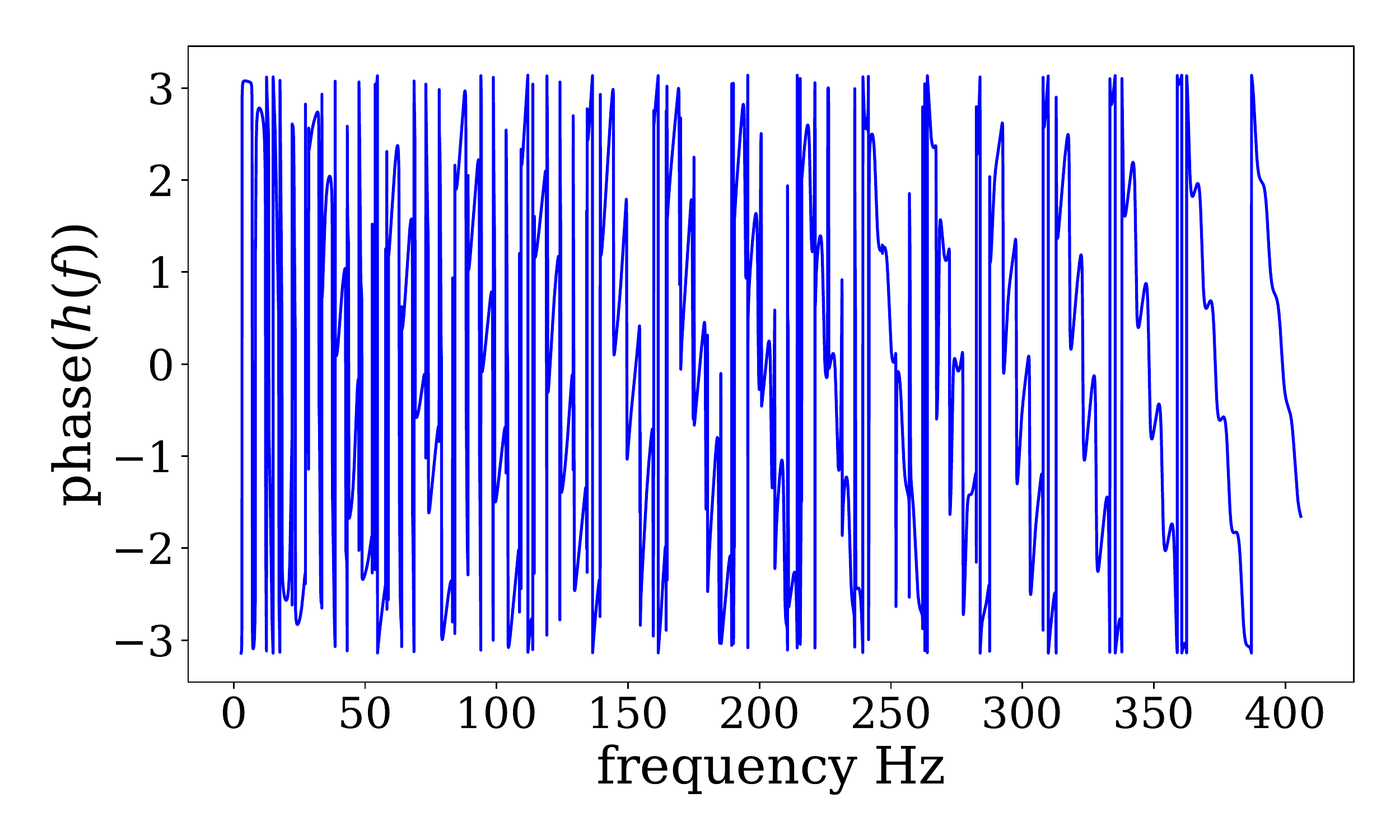}  
\caption{a) $|h(f)|$ as reconstructed from Re$(h(t))$ b) The phase of $h(f)$.}
\label{fig2}
\end{figure}

We show the resulting $|h(f)|$ in Fig.~\ref{fig2}a with $N_E=180$, $M=50M_\odot$. The range of frequencies over which the resonance spectrum extends is proportional to $1/M$, and will thus be different for every event. $N_E$ along with the choice of $R_{\rm wall}$, for this example chosen to be $-0.995$, controls the overall height of the spikes. Relative to a fixed amount of noise, this overall height will increase with $N_E$ up to roughly $N_E\sim200$, after which the highest set of peaks, the ones that contribute most to a signal, begin to shrink. This behavior is controlled by the value of $R_{\rm wall}$ and it occurs even though the energy radiated per time delay, for instance, is a monotonically decreasing function of time.

In Fig.~\ref{fig2}a we see the two component structure of $|h(f)|$, which arises due to the real projection. The lower component originates from what was the negative frequency part of the original $\psi_\omega$. The relative size of the two components is controlled by the function $f(\omega)$, which as introduced above emphasizes positive frequencies. On the other hand, the source integral $|D(\omega)|$ contains the factor $|\omega-\omega_0|$ that enhances the original negative frequency part \cite{Conklin:2019fcs}. Thus in our example the negative frequency component in $|h(f)|$ is still comparable in size to the positive frequency component.

Other echo searches have tended to try to model the full waveform. In frequency space the full information is in $|h(f)|$ and the phase of $h(f)$. The latter is displayed in Fig.~\ref{fig2}b. In a standard matched-filter search, the templates must be constructed to match this information as well. This makes clear why such searches are liable to fail. For a resonance search, only $|h(f)|$ is needed.

\begin{figure}[t]
   \includegraphics[width=0.48\textwidth]{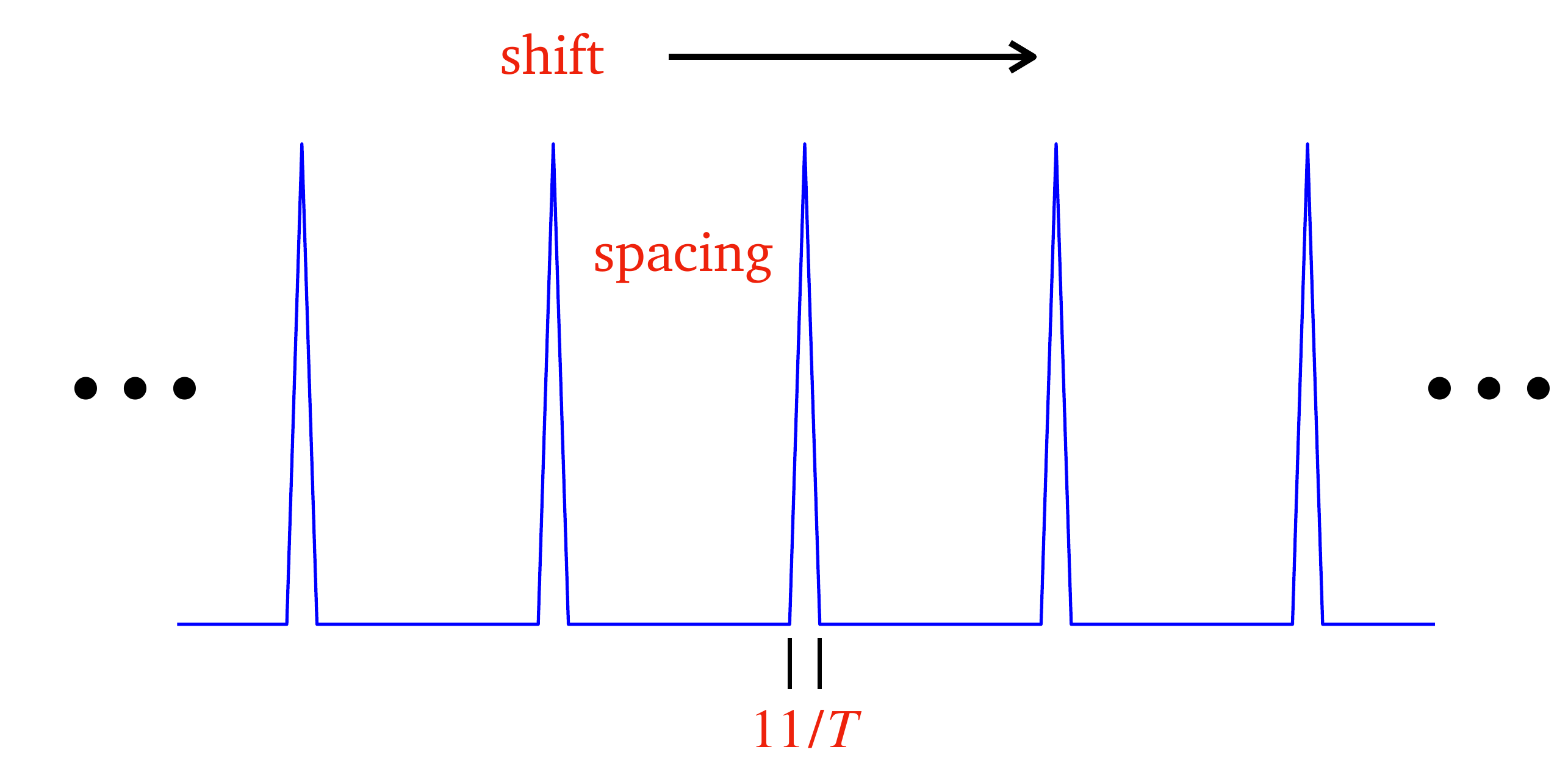}
	 \includegraphics[width=0.48\textwidth]{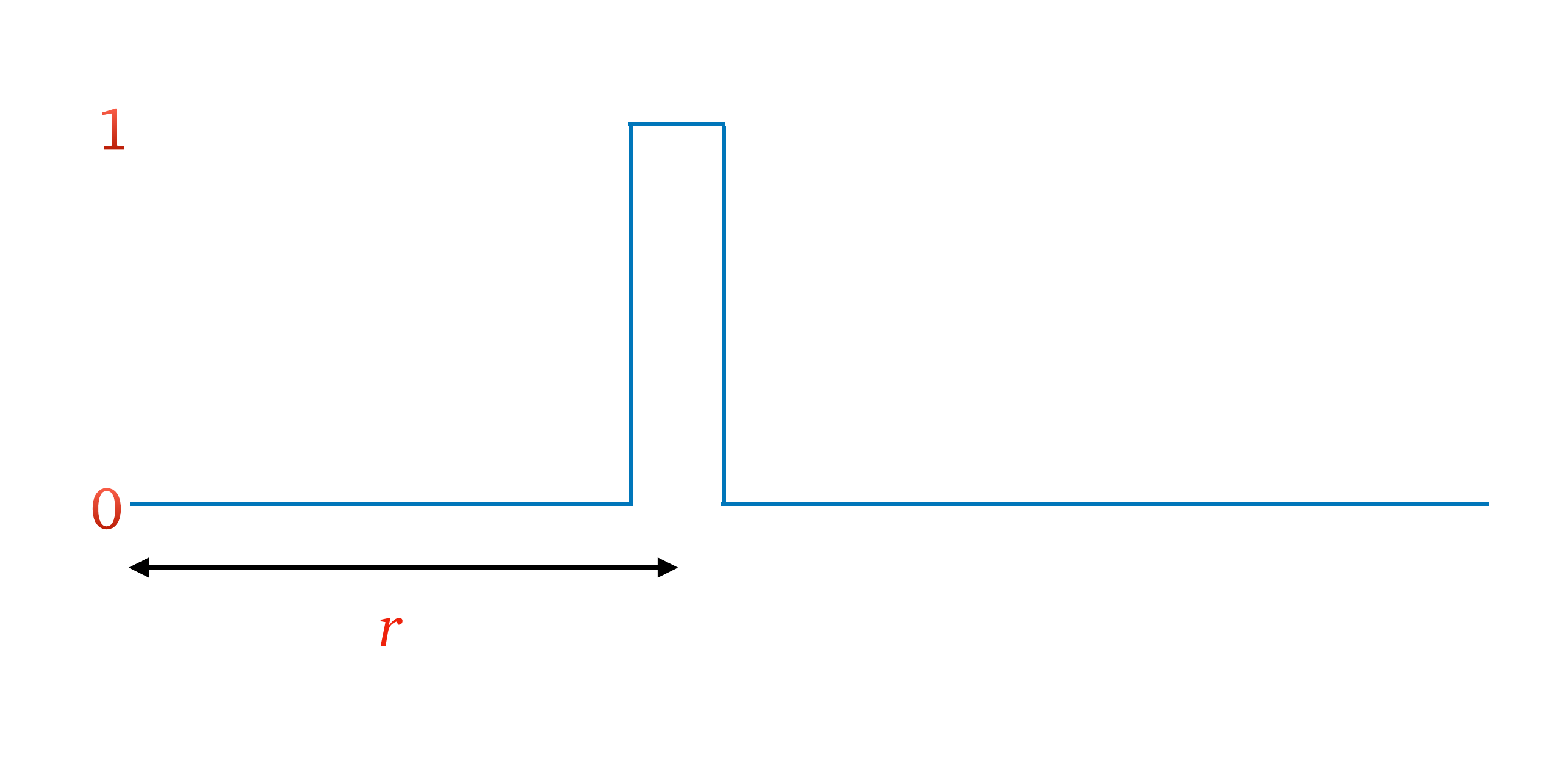}
\caption{a) Five teeth of a uniform comb in frequency space. b) The vector $V(r)$ of length $N_E$ is composed of 23 1's with location shown and the rest 0's.}
\label{fig3}
\end{figure}

The construction $|\textrm{FFT(whitened data of duration }T\textrm{ after merger)}|$ is a data series in the frequency domain. Onto this we apply a frequency bandpass $f_{\rm min}<f<f_{\rm max}$ to model the finite extent of the resonance spectrum. Then we multiply by a comb structure characterized by a spacing between teeth and an overall shift. The comb is uniform, meaning all teeth are the same with constant spacing, as shown in Fig.~\ref{fig3}a.\footnote{This is a simplification of our previous analysis \cite{Conklin:2017lwb}.} We thus first define a vector that is the result of multiplying the comb onto the data for a fixed comb spacing,
\begin{align}
U={\rm mean}\left[|{\rm FFT}(\mathit{data})|\times {\rm comb}(\mathit{shift})\right]
,\end{align}
where the mean is with respect to the nonzero values. $U$ is a vector in shift space with length $N_E$. A signal and the correct comb spacing will result in larger values around some shift. Let $V(r)$ be another vector that represents an idealized bump centered at a shift of $r$. This is illustrated in Fig.~\ref{fig3}b. Now we can take the two vectors $U_H$ and $U_L$, from the Hanford and Livingston detectors respectively, and form the following quantity,
\begin{align}
C=\max_r\left[\textrm{Correlation}(U_H,V(r))\times\textrm{Correlation}(U_L,V(r))\right]
.\end{align}
$C$ is our correlation \cite{Conklin:2017lwb} between detectors and it is still a function of the comb spacing $\Delta f=1/\Delta t$. A signal shows up as a peak in a plot of $C(\Delta f)$. This will be a sharp peak because the spacing needs to be tuned to achieve an overlap with the large number of sharp spikes.

The whitening of the original strain data is an important first step in the analysis, and to obtain the power spectral density (PSD) required for whitening, an averaging over 300 segments is performed.\footnote{We use the Welch method with the Hanning window and with some overlap and padding turned on. The same settings are used for all events.} The range of data used for obtaining the PSD is larger than the search region, and it is centered on the midpoint of the search region. The duration of the segments (in seconds) used in each event is given in Table \ref{a1} below. Allowing a binary choice of 0.5 or 1 s seems to better treat the different noise characteristics in different events.\footnote{Incomplete data for GW151226 prevented the choice of 1 s.} A trade-off is involved since although longer segment times will reduce noise on finer-grained frequency scales, it increases the risk of altering the signal that also occurs on fine-grained scales. One further noise cut is applied ro the frequency series $|\textrm{FFT(whitened data of duration }T\textrm{ after merger)}|$. Any value in this series that is more than 3 times the average value is reduced to the 3 times value. Such fluctuations should only occur about 0.1\% of the time for Gaussian noise, but in the earlier events especially, there is substantial non-Gaussian noise. It is fair to assume that the signal spikes are typically not high enough to be much affected by this cut.

\section{Results}\label{s3}

For a truncated Kerr black hole, the time delay $\Delta t$ has known dependence on mass $M$, spin $\chi$ and redshift $z$ \cite{Abedi:2016hgu,Cardoso:2017njb},
\begin{align}
\frac{\Delta t}{M}=4\,{\color{red}\eta} \log(\frac{M}{\ell_{\rm Pl}})\;(\frac{1+(1-\chi^2)^{-\frac{1}{2}}}{2})\;(1+z).
\label{e1}\end{align}
The redshift factor is due to $\Delta t$ and $M$ being measured in the detector and source frames respectively. $\Delta t/M$ is driven large by the large log, due to the presence of the Planck scale. We have introduced the parameter $\eta$ to characterize the small distance $\delta r$ from the would-be horizon out to where strong gravity extends, as follows,
\begin{align}
\delta r\approx \left(\frac{\ell_{\rm Pl}}{M}\right)^{({\color{red}\eta}-1)}\ell_{\rm Pl}\approx\left(\frac{M}{\ell_{\rm Pl}}\right)^{(2-{\color{red}\eta})}\bar\ell_{\rm Pl},
\end{align}
where $\bar\ell_{\rm Pl}$ is the proper Planck length. Thus if $\eta$ is between 1 and 2, then $\delta r$ is smaller than the coordinate Planck length $\ell_{\rm Pl}$ and is larger than the proper Planck length. Because of the many evenly spaced spikes, a resonance signal will determine $\Delta t$ with negligible error. The LIGO measurements of $M$, $\chi$ and $z$ for the final BH have more significant errors, and by ignoring correlations in these errors for simplicity, we can thus determine an $\eta$ and its error from a measurement of $\Delta t$.

\begin{figure}[h]
\centering
   \includegraphics[width=0.7\textwidth]{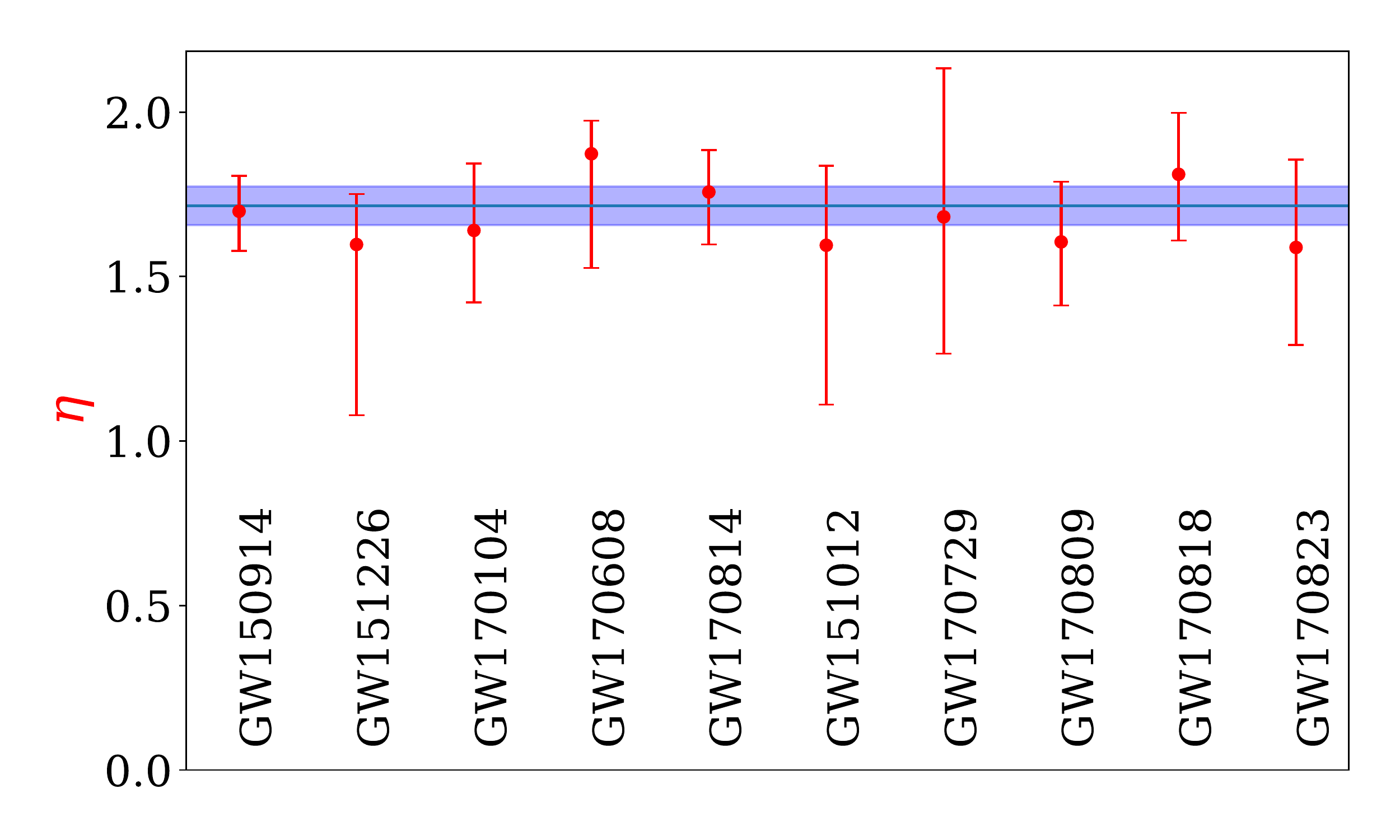}
\caption{Determination of $\eta$ using (\ref{e1}).}
\label{fig4}
\end{figure}

We find resonance signals for all events. Before elaborating on this we first provide a summary plot in Fig.~\ref{fig4}, where we show our determination of $\eta$ from all ten BH merger events. The consistency for a common value of $\eta$ describing all events is excellent. The value $\eta=1.72\pm0.06$ means that $\delta r\approx 10^{-28}\ell_{\rm Pl}\approx10^{12}\bar\ell_{\rm Pl}$. We stress that the size of the error bars has nothing to do with the strength of the resonance signal for each event. We shall address the question of signal strengths below.

\begin{figure}[h]
\centering
   \includegraphics[width=0.48\textwidth]{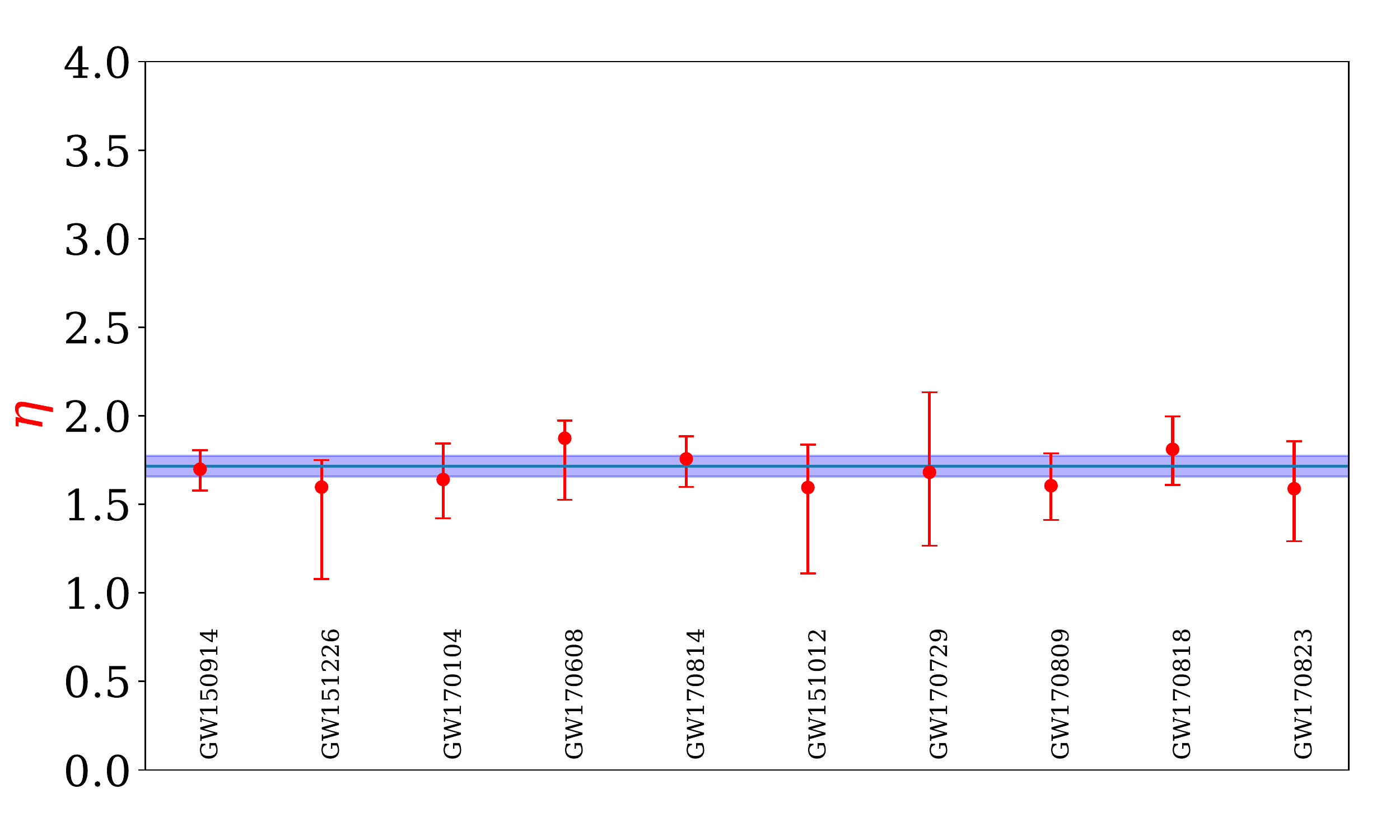}
   \includegraphics[width=0.48\textwidth]{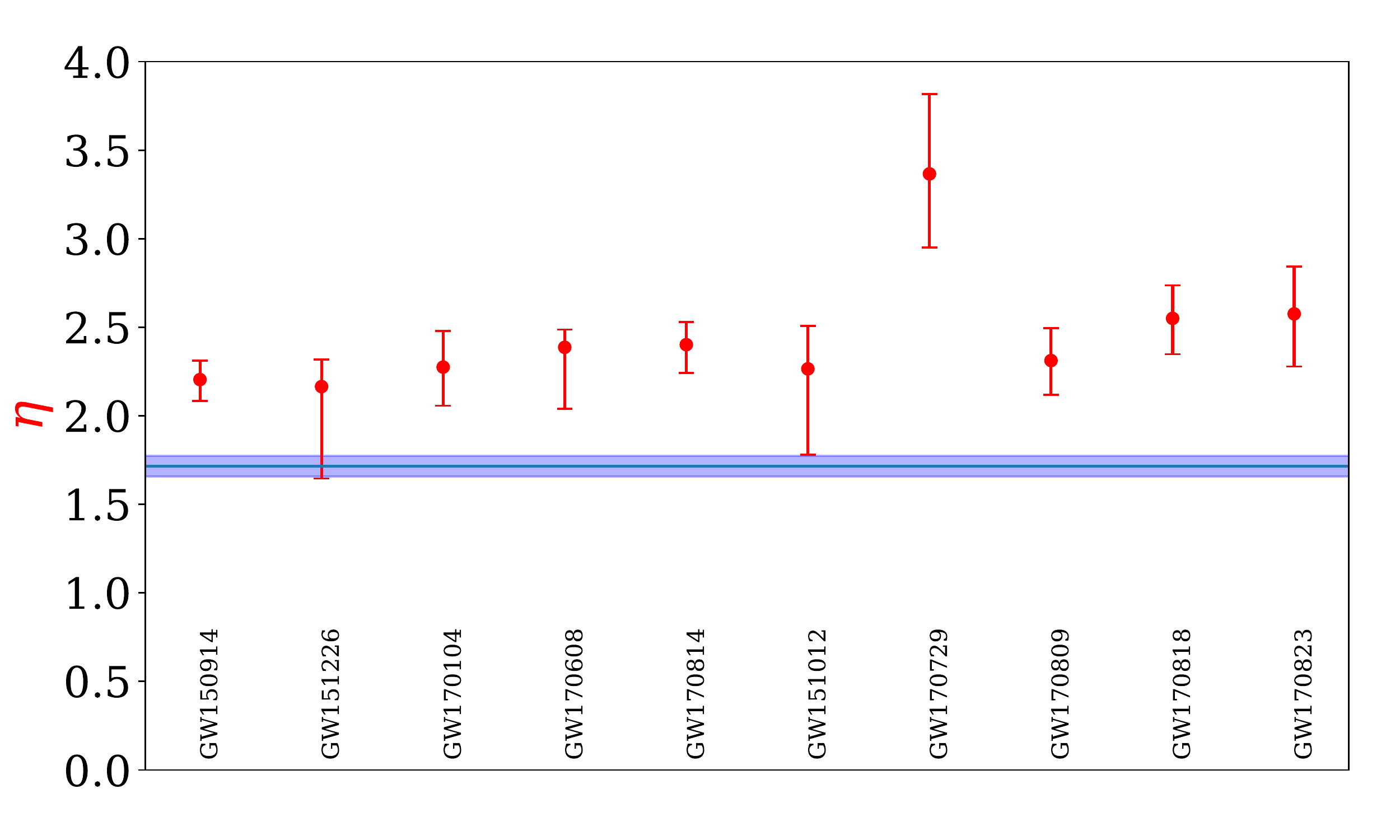} 
\caption{Left: Same as Fig.~\ref{fig4}. Right: Removal of spin and redshift factors in (\ref{e1}).}
\label{fig5}
\end{figure}

These results provide a good test of the spin and redshift dependence in (\ref{e1}). This is shown more clearly in Fig.~\ref{fig5}, where the left plot is the same as Fig.~\ref{fig4} but with an enlarged range of $\eta$, while the right plot shows the result when removing the spin and redshift dependent factors from (\ref{e1}). The indication of a common value for $\eta$ is lost, and the individual values of $\eta$ rise above 2. Values this large would correspond to $\delta r$ values that are much smaller than the proper Planck length, which does not seem reasonable. Thus the data is already supporting the truncated Kerr BH model in (\ref{e1}). Fig.~\ref{fig5} also draws attention to event GW170729, for which the three quantities, $M$, $\chi$ and $z$, are all significantly larger than for the other events.

We now take a closer look at the analysis and the results for each event. For each event we will show the predicted spectrum $|h(f)|$. To simplify the generation of these spectra we assign one of three spin values to each event. These values are well within the spin uncertainties, and so we group the ten events as in Table \ref{a1}. There we also give our chosen values of parameters $R_{\rm wall}$ and $\kappa$ for these three groups of events, which along with $M$, $\chi$, $\Delta t$ and $N_E$ are used to obtain $|h(f)|$.

\renewcommand{\arraystretch}{1.3}
\renewcommand{\tabcolsep}{4mm}
\begin{table}[h]
\begin{tabular}{|c|c|c|c|}
\hline
Spin ($\chi$)&2/3&0.72&0.81\\\hline
&GW150914 (.5)&GW151226 (.8)&GW170729 (.5)\\\cline{2-4}
&GW170104 (1)&GW170814 (1)& \\\cline{2-4}
&GW170608 (1)&GW170809 (1)&\\\cline{2-4}
&GW151012 (.5)&GW170823 (.5)&\\\cline{2-4}
&GW170818 (.5)&&\\\hline
$R_{\rm wall}$ & $-0.995$& $-0.994$& $-0.992$\\\hline
$\kappa$&13&10&2 \\\hline
\end{tabular}
\caption{The ten events are grouped according to spin, along with other chosen parameters. The numbers beside event names are segment durations in seconds used in whitening.}\label{a1}
\end{table}

Except for the binary choice of a whitening parameter, as indicated in Table \ref{a1}, the analysis carried out for each event is exactly the same. There are no other procedures or parameters that vary between events. We will express the comb spacing $\Delta f$ as an integer, $n=\Delta f/(1/T)=T/\Delta t$. It is with respect to $n$ that we determine the correlation between detectors as described earlier. As we have said, a signal peak should be very sensitive to choice of $\Delta t$. In the process of determining the true $\Delta t$, we can choose $T$ such that the peak location is fixed at a particular integer value of $n=N_E$, that is that $T$ is $N_E$ times the true $\Delta t$.

We then further optimize the signal (the peak height) by varying three quantities: $N_E$ and the upper and lower limits of the bandpass $f_{\rm min}$ and $f_{\rm max}$. For each event the optimal bandpass is displayed as a band on a plot. We also give the optimal $N_E$ (as a multiple of ten) and the measured values of $\Delta f$ and $\Delta t/M$ (where the uncertainty inherent in $M$ is not displayed). We have already seen that it is the spin and redshift factors that account for the different values of $\Delta t/M$ from different events. The plot will compare the optimal bandpass to the resonance pattern in $|h(f)|$. We stress that the optimal bandpass is determined solely from the data, while the resonance pattern is a theoretical quantity. It is interesting to note the consistency between the bandpass and the resonance pattern across all events.

We also wish to give a sense of the size of the signal relative to noise, with the latter being the square root of the PSD, that is the amplitude spectral density (ASD), for each detector (as was used in the whitening). To make a comparison of the signal $|h(f)|$ with the noise possible, we shall effectively calculate $|h(f)|$ also as an ASD, but without employing the averaging and windowing (i.e.~the Welsh method) that was used to calculate the noise ASD. This is equivalent to producing $|h(f)|$ through a FFT, but in a way that gives the proper normalization to allow comparison of signal and noise. 

To help pin down the strength of the signal we inject our time-domain waveform Re$(h(t))$ into some stretch of strain data from each detector, away from the signal region. By running this signal plus noise through our analysis we find that the resulting peak on the injected signal plot can be similar to that on the actual signal plot. This then provides a rough estimate of the required strength of the signal, and we incorporate this into the normalization of $|h(f)|$ on the plot. We also provide the corresponding peak value of $|h(t)|$, where the peak value occurs in the first echo.\footnote{Our estimate of the peak value assumes the simplest initial perturbation, that of a single pulse moving back and forth in the cavity. For a more general perturbation, the required peak value of $|h(t)|$ may be substantially less.} We do not include the first pulse, that is the pulse before the first echo, in our injected waveform as this pulse is to be associated with the actual merger and ringdown. The starting time of our signal region is always taken to be after the ringdown has occurred. Our plots show separately the small ASD of the first pulse as a black line. 

We also show a second plot for each event that shows the distribution of background values of the correlation along with the size of the correlation at the signal peak. We will discuss this more in the next section on p-values.

These plots and results are presented below for the five spin 2/3 events first (Figs.~6-10) and then the four spin 0.72 events (Figs.~11-14). Each figure shows the spectral plot and the correlation plot. The remaining event GW170729 (Fig.~15) has been flagged above, and we find that its results are both interesting and different. Here we find two resonance signals, for two nonoverlapping choices of the bandpass. Both of these signals are strong, as seen from the two correlation plots. Each signal produces its own value of $\Delta f$ that happens to be essentially the same. From the displayed $|h(f)|$ we see that the high spin causes the positive and negative frequency components to become more separated, and the two signals are consistent with observing these two components separately. When attempting to use a broader bandpass that would encompass both components, the signal effectively disappears. Given that each signal has the same resonance spacing, the implication is that the two resonance patterns are displaced relative to each other, in such a way that the broad uniform comb misses one or the other. Such a relative shift is natural and it depends sensitively on $\omega_0$, which pins the overall shift of $\psi_\omega$. Since the spin and thus $\omega_0$ is not measured accurately enough, this shift cannot be predicted.

Finally we present results for the neutron star merger event, GW170817 in Fig.~16. We reported a resonance signal for this event in \cite{Conklin:2017lwb}, but now we pass it through the current analysis. The results remain similar to before. The time scales are shorter for this event and the high sampling rate data (16384 Hz) is used. We use 200 segments of 1/4 s duration for the whitening. LIGO does not provide the final mass and spin, so here we make choices ($M=2.2M_\odot$ and $\chi=2/3$) that give a sensible looking $|h(f)|$ when compared to the optimal bandpass. The measured $\Delta t$ then corresponds to $\eta=1.6$, and increasing $M$ would decrease $\eta$ further. We do not use this event in the analysis of the next section.

\section{P-values}
In this section we shall obtain probabilities useful for estimating p-values. By using many stretches of data we can quite accurately determine the background distribution of our correlation quantity, separately for each event. Then we can find the probability of finding a value at least as high as our signal peak. A p-value of each event should also account for the fact that the measured $\Delta t$ gives an $\eta$ that is consistent with the nine other events, as per Fig.~\ref{fig4}. So for example if we require that $\eta$ be within $2\sigma$ of its central value, then this gives the range of $\Delta t$ (or a range of $n$ on a signal plot) over which we can look for a background correlation as high as our signal correlation.

We use data throughout the 4096s segment that contains the merger event, and within this we use segments of the same length as in the signal analysis with locations chosen randomly for each detector. The whitening of the data is repeated and the already determined values of $\Delta t$, $f_{\rm min}$, $f_{\rm max}$ and $N_E$ are used to produce many background versions of the signal plot. Each such plot has $n$ ranging from 0.6 to 1.4 times $N_E$. This procedure has an important benefit, namely that we can verify that the correlation at the central value, where the peak occurs in the signal plot, does not have any excess strength in the background plots.\footnote{The one exception is GW150914 where there is some excess correlation at or close to the central value. This may be related to how close $\Delta f=3.991$ Hz is to 4 Hz, thus suggesting that there is some 4 Hz correlation generated by noise. This excess correlation is weaker than the signal.} A correlation that exists outside the signal region would indicate that it is being generated by noise.

By collecting together all the correlation values from all the background plots, we can then determine their distribution. We find that a good fit is obtained to a function of the form $x^\alpha e^{-\beta x}$, which when normalized is a generalized gamma probability density function $P_{\rm gg}(x)$.  The fit (different for each event) has four parameters, $\alpha$, $\beta$, and a scale and a location parameter. These parameters are determined by a maximum likelihood estimate directly from the correlation values ($\sim 20000$ values). Typically $\beta$ is between 0.5 and 1 and the location parameter is close to vanishing.

A plot for each event shows $P_{\rm gg}(x)$ (red curve) for the correlations from background. The histogram is added to indicate how well $P_{\rm gg}(x)$ fits the data. Also shown are the signal peak values of the correlation. The signal peak value $x_{\rm sig}$ determines the probability $\int_{x_{\rm sig}}^\infty dx P_{\rm gg}(x)$ that any single background measurement can produce a value of $x$ at least as high. We then multiply by the number values of $n$ that lie within $2\sigma$ of $n=N_E$ (explicitly, this number of values is $0.14N_E$). The resulting numbers are simple estimates of p-values and they are shown in Table \ref{a2}.

\begin{table}[h]
\begin{tabular}{|c|c|c|c|}
\hline
GW150914&0.008 &GW151226 &0.014 \\\hline
GW170104&0.33 &GW170814 &0.098 \\\hline
GW170608&0.038 &GW170809 &0.081\\\hline
GW151012&0.0016 &GW170823 &0.026\\\hline
GW170818&0.0094 &GW170729 &0.0010\;\&\;0.0006\\\hline
\end{tabular}
\caption{Simple p-values for the ten black hole merger events. These are the probabilities that background noise can generate the signal peak height or higher for any value of $n$ within 7\% (2$\sigma$) of $N_E$.}\label{a2}
\end{table}

These simple p-values have the advantage of being well defined and they usefully show the relative signal strength of the different events. They highlight again the significance of the GW170729 results. From the individual p-values we can contemplate a global p-value. Combining p-values is not a completely well-defined procedure and various methods are used. Combining the 11 p-values in Table \ref{a2} according to the Fisher and Stouffer methods\footnote{The Stouffer method may be the less well known, but it is easy to describe: convert each p-value $p_i$ to $x_i$ using $p_i=\int_{x_i}^\infty dx P_{\rm norm}(x)$, get the combined $x_c=\sum_i x_i/\sqrt{N}$, and then convert $x_c$ back to the combined p-value $p_c$.} (which assume independence of the p-values to be combined) gives $5\times10^{-11}$ and $2\times10^{-12}$ respectively.

We could expect that the individual p-values become somewhat less trustworthy as the signal value $x_{\rm sig}$ goes further out onto the tail of the distribution, since the analytical $P_{\rm gg}(x)$ is being used to extrapolate into this region where there are few background values. Thus the lowest p-values Table \ref{a2} may have relatively larger uncertainty. To consider the impact of this on the combined p-value, we can for example add a fixed amount to \textit{all} the p-values, which relatively impacts the lowest ones the most. For example when we add 0.005, the combined p-values become $1\times10^{-8}$ and $2\times10^{-10}$. This is just for illustration, and 0.005 seems far more than needed to account for the PDF uncertainty.

In contrast to our simple p-values, we could consider obtaining p-values by repeating our \textit{entire} analysis, many times, on background data. Our search would have to be automated and be applicable to both signal and background regions. p-values may depend on how this is accomplished. Although we do not accomplish this here, it would entail specifying the ranges of quantities, over which they are varied to find the signals. This includes $\Delta t$ itself as well as the frequency bandpass limits $f_{\rm min}$ and $f_{\rm max}$ and the integer $N_E/10$. We could then find the probability for finding a background correlation as high as our signal correlation, for each event, and thus a set of $N$ p-values (here $N=11$). The combined p-value would then be the probability of observing correlations as extreme as our signal correlations, over the ranges specified and assuming background only.

But there is more to our signals than the correlation size. For a particular background that can generate a signal-size correlation, the above-mentioned quantities now take whatever random values the background produces. One then needs to consider the probability that these random values could display the regularities actually observed in the signals. For $\Delta t$ we have taken the regularity to be the $2\sigma$ version of the narrow band in Fig.~\ref{fig4}. If this corresponds to a fraction $p_1$ of the $\Delta t$ range actually explored for each event, then the probability would be $\sim p_1^N$. For the bandpass, it is the probability that each of $f_{\rm min}$ and $f_{\rm max}$ lies in the range that is compatible with our theoretical spectra, as happens for the signals. We write this probability as $\sim p_2^Np_3^N$. For $N_E$, it is the probability that each event has correlation peaks for a range of $N_E$ (even though we have chosen only one in each case) and that these ranges largely overlap between events, as happens for the signals (other than GW151226). We write this probability as $\sim p_4^{N-1}$. The final probability of interest would then be the combined p-value times $\sim(p_1p_2p_3)^Np_4^{N-1}$. In fact we could consider increasing the search ranges until the combined p-value is no longer very small. But then the $p_i$'s will be small, and we see that the $p_i$'s do not have to be very small to imply by themselves a very small final probability.

The point is that in addition to evidence against the background hypothesis, we are seeing evidence for the signal hypothesis. The signal hypothesis involves a simple theoretical model that is predicting how the various quantities in each event should be related, and the data is largely conforming to these predictions. Returning to our simple p-values, they are being defined for each event by holding $f_{\rm min}$, $f_{\rm max}$ and $N_E$ fixed to their values from the signal region. We are ignoring both the search ranges of these quantities and how the observed regularities in these quantities support the signal hypothesis. These are compensating effects and so our simple p-values are a sensible first step to assess the significance of our results. Our simple p-values also ignore the additional evidence for the signal hypothesis coming from GW170729, namely that the two different bandpasses have produced the same $\Delta f$.

\section{Conclusions}

LIGO is sensitive to the cavity resonance structure of not quite black holes. This particular sensitivity to Planck-scale physics is underappreciated. The full signal waveform produced by a not quite black hole may have too much information and model dependence for a standard matched-filter analysis. On the other hand the evenly spaced resonance pattern is robust and relatively easy to search for. This is the approach taken here.

Evidence is already accumulating from the ten BH merger events reported in \cite{LIGOScientific:2018mvr}. We have described the consistency of the measured $\Delta t$ values with the mass, spin and redshift dependence predicted from the simple truncated Kerr BH model. Event GW170729 in particular is sufficiently different from the other events to provide a good test of the model. We have highlighted the strength of its signal and the way it may already be showing evidence of the expected two component structure of the resonance pattern \cite{Conklin:2019fcs}. We have also displayed the consistency of the data-driven bandpasses with the spectrum predicted for each event in the same model. The measured $\Delta t$ values point to a common distance scale $\delta r$, with our measured value being larger than the proper Planck length, $\delta r\approx10^{12}\bar\ell_{\rm Pl}$.

The LIGO data certainly displays non-Gaussian noise, some of it not understood, and some of it displaying comblike features reminiscent of our signal. But given the observed regularities of the signals, and in addition the low p-values, it is now very unreasonable to expect that this could explain our signals. Our simple p-values are both transparent and able to provide a sensible first estimate of the significance of our results. The combined p-value is found to be very small.

We have stressed that our search strategy is uniform across all events. But to define the strategy, some choices and parameters were fixed once and for all. With the strategy now frozen, it will be important to apply it to new data. Many new black hole merger events have already been discovered in the LIGO-Virgo O3 run and it only remains to look at this data. We have one hope concerning the data. In their continuing efforts to reduce noise with respect to their merger-signal target, it is to be hoped that LIGO-Virgo does not inadvertently remove poorly understood non-Gaussian noise that, as seen here, could harbor new physics.\footnote{In this connection we note that a strong resonance signal for event GW170104 as described in \cite{Conklin:2017lwb} was based on an early O2 data release. This signal weakens considerably when the subsequent GWTC-1 data release \cite{LIGOScientific:2018mvr} is used instead, and now GW170104 gives our weakest signal. For all events in the present analysis we use only the GWTC-1 data release.}

\newpage
\begin{table}[h!]
\raisebox{15ex}{{FIG.~6.~GW150914}}\hspace{5ex}\begin{minipage}[b]{0.58\linewidth}
\centering
\includegraphics[width=\textwidth]{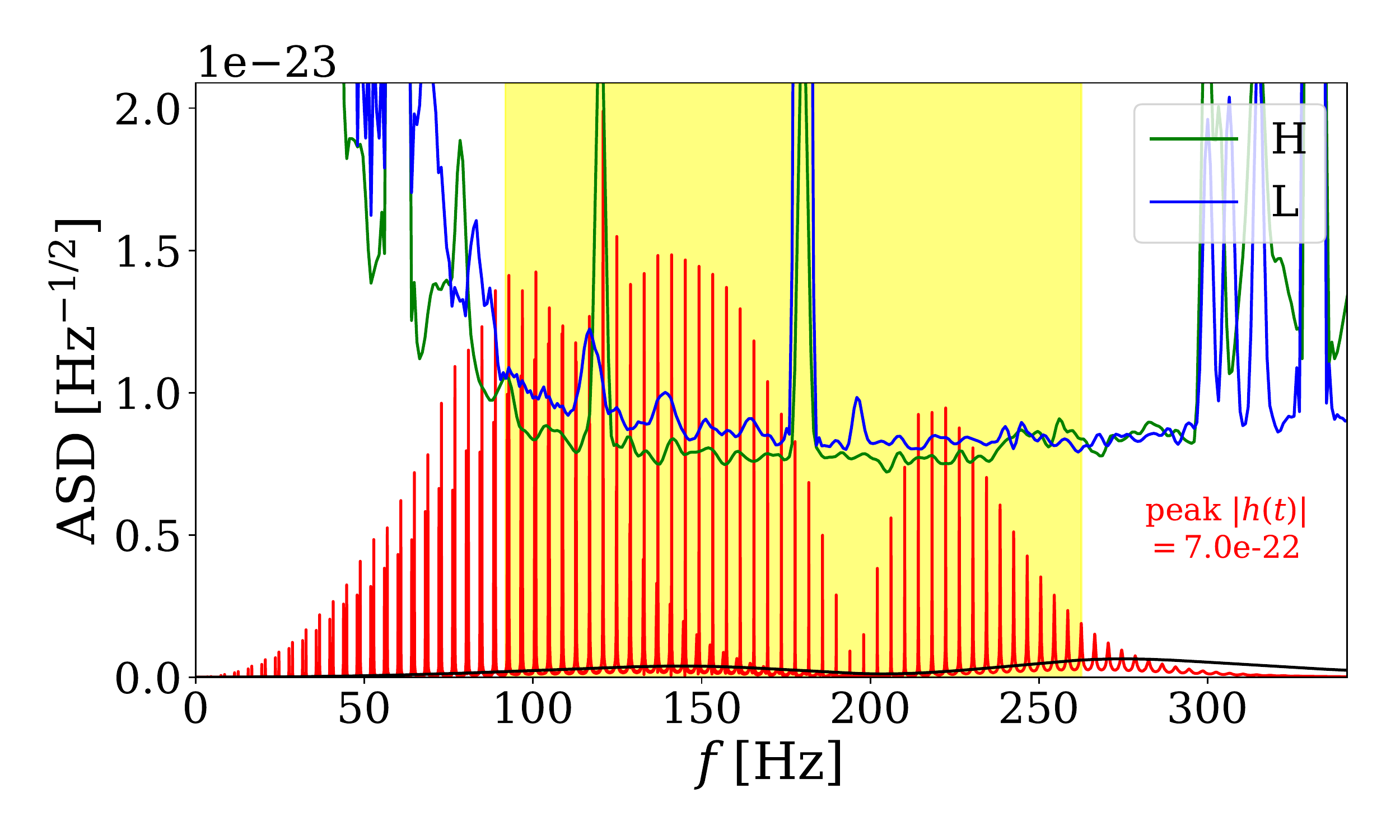}
\end{minipage}\\
\raisebox{15ex}{\begin{minipage}[b]{0.2\linewidth}\centering
\renewcommand{\arraystretch}{1.5}\small
$\begin{array}{|c|c|}
\hline
N_E&200\\
\hline
\Delta f&3.991{\rm Hz}\\
\hline
\displaystyle\frac{\Delta t}{M}&806\\[5pt]
\hline
\chi&2/3\\
\hline
\end{array}$
\end{minipage}}
\hspace{0.5cm}
\begin{minipage}[b]{0.52\linewidth}
\centering
\includegraphics[width=\textwidth]{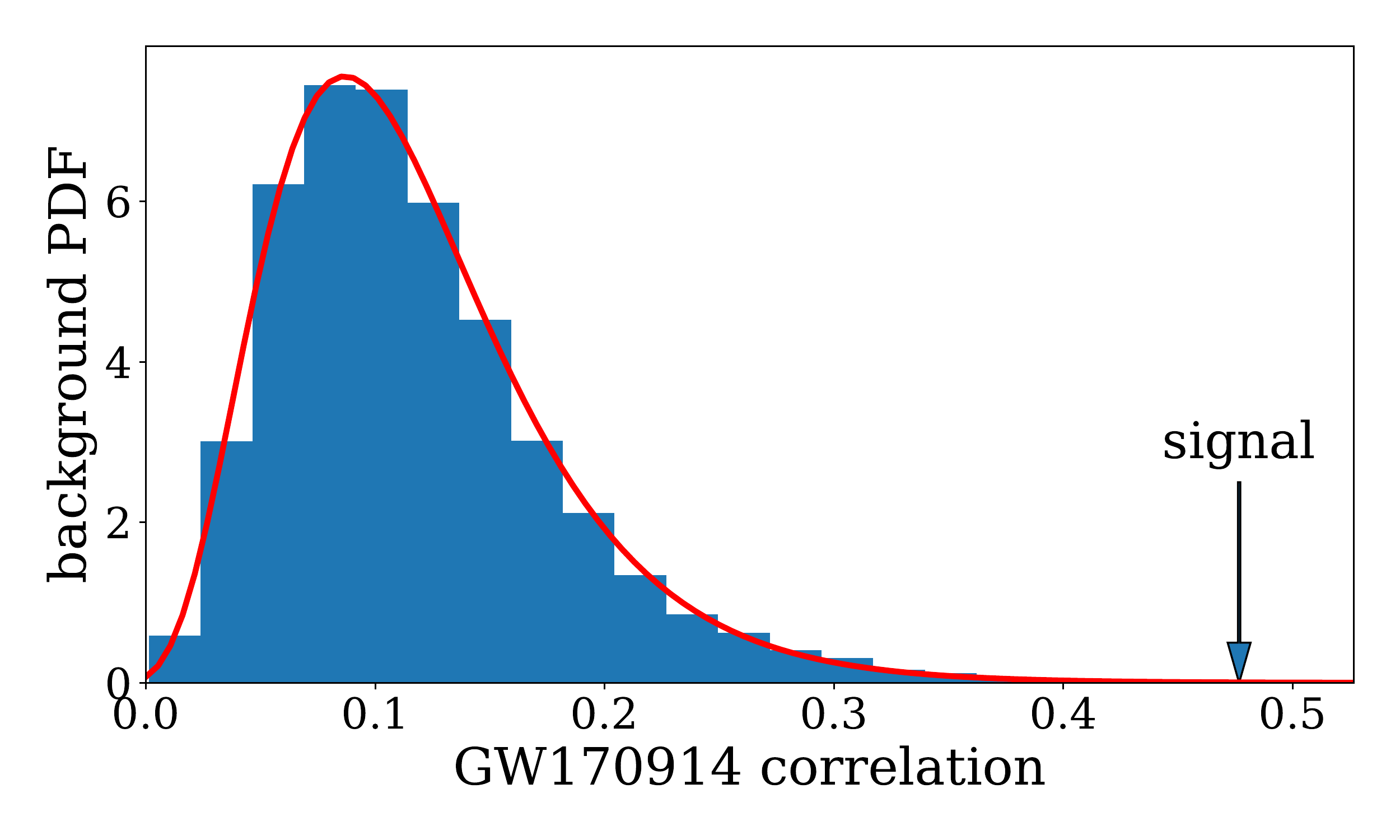}
\end{minipage}
\end{table}
\begin{table}[h!]
\raisebox{15ex}{{FIG.~7.~GW170104}}\hspace{5ex}\begin{minipage}[b]{0.58\linewidth}
\centering
\includegraphics[width=\textwidth]{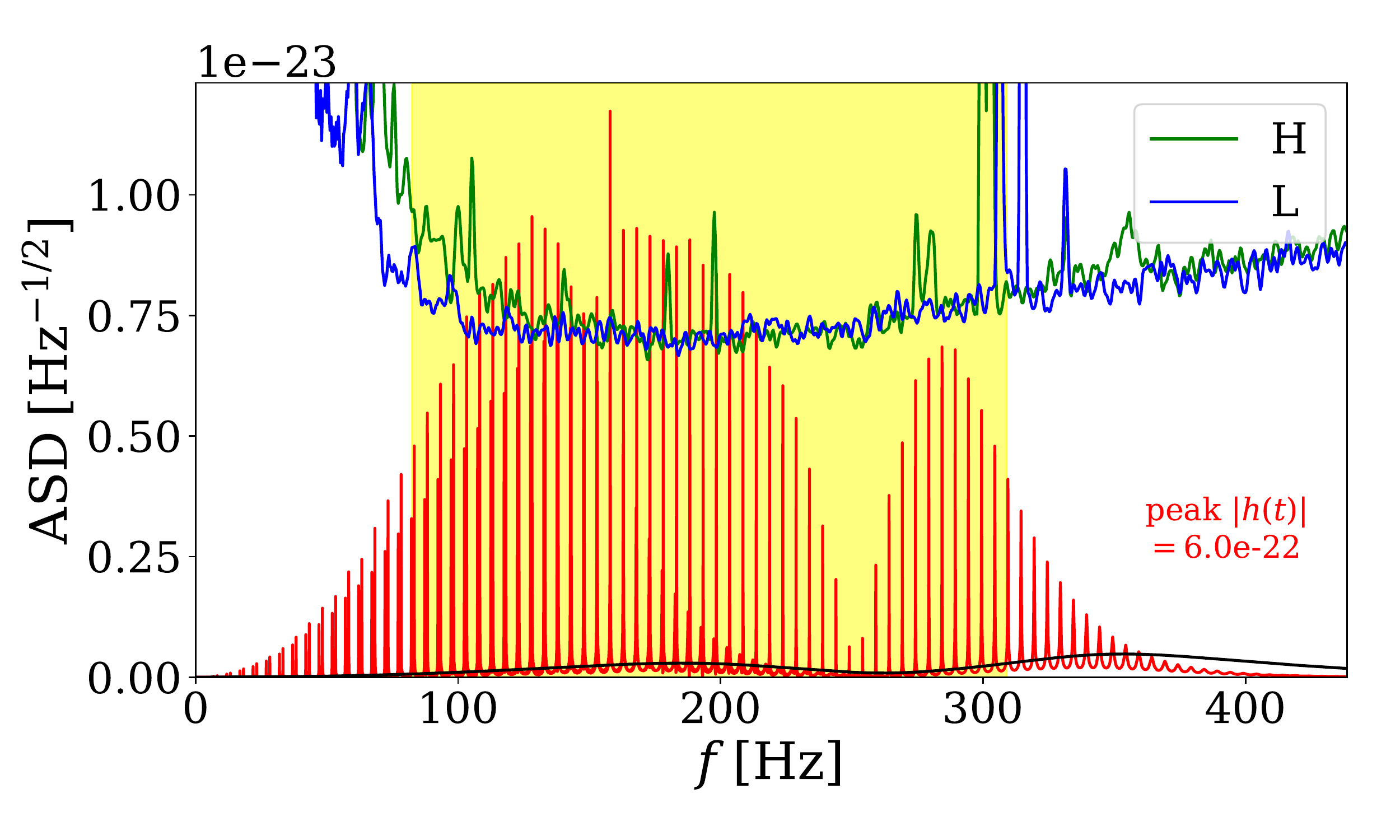}
\end{minipage}\\
\raisebox{15ex}{\begin{minipage}[b]{0.2\linewidth}\centering
\renewcommand{\arraystretch}{1.5}\small
$\begin{array}{|c|c|}
\hline
N_E&150\\
\hline
\Delta f&4.976{\rm Hz}\\
\hline
\displaystyle\frac{\Delta t}{M}&831\\[5pt]
\hline
\chi&2/3\\
\hline
\end{array}$
\end{minipage}}
\hspace{0.5cm}
\begin{minipage}[b]{0.52\linewidth}
\centering
\includegraphics[width=\textwidth]{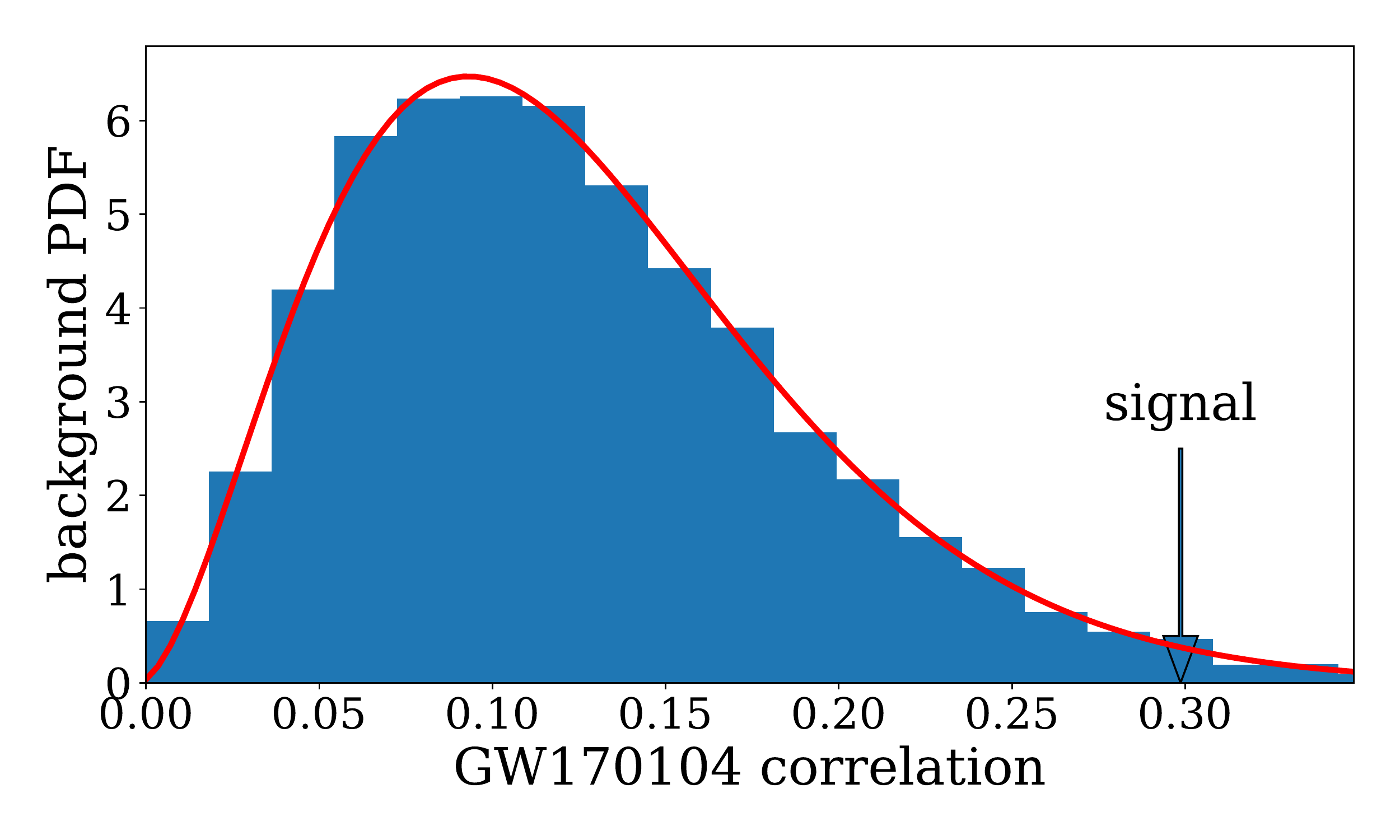}
\end{minipage}
\end{table}

\begin{table}[h!]
\raisebox{15ex}{{FIG.~8.~GW170608}}\hspace{5ex}\begin{minipage}[b]{0.58\linewidth}
\centering
\includegraphics[width=\textwidth]{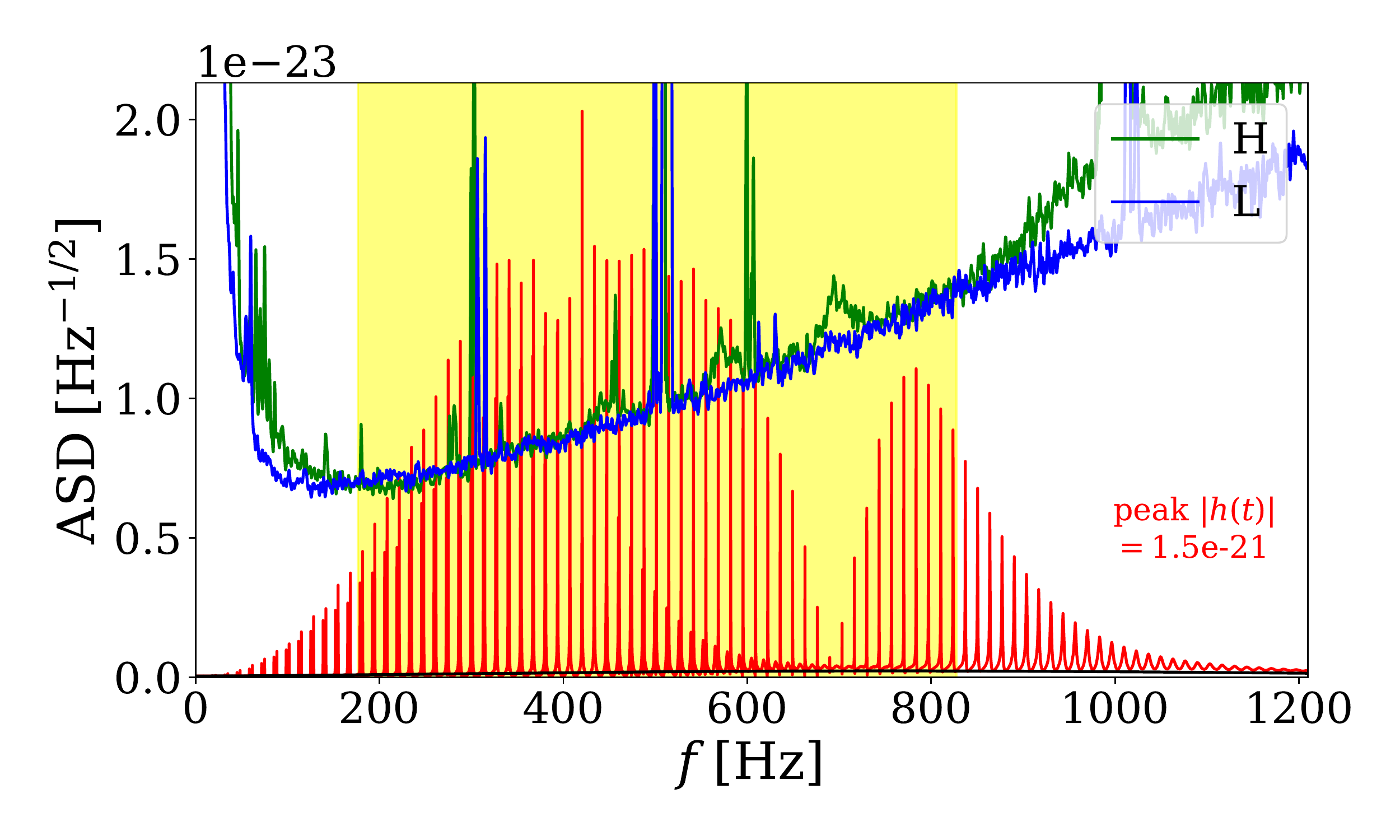}
\end{minipage}\\
\raisebox{15ex}{\begin{minipage}[b]{0.2\linewidth}\centering
\renewcommand{\arraystretch}{1.5}\small
$\begin{array}{|c|c|}
\hline
N_E&200\\
\hline
\Delta f&13.226{\rm Hz}\\
\hline
\displaystyle\frac{\Delta t}{M}&862\\[5pt]
\hline
\chi&2/3\\
\hline
\end{array}$
\end{minipage}}
\hspace{0.5cm}
\begin{minipage}[b]{0.52\linewidth}
\centering
\includegraphics[width=\textwidth]{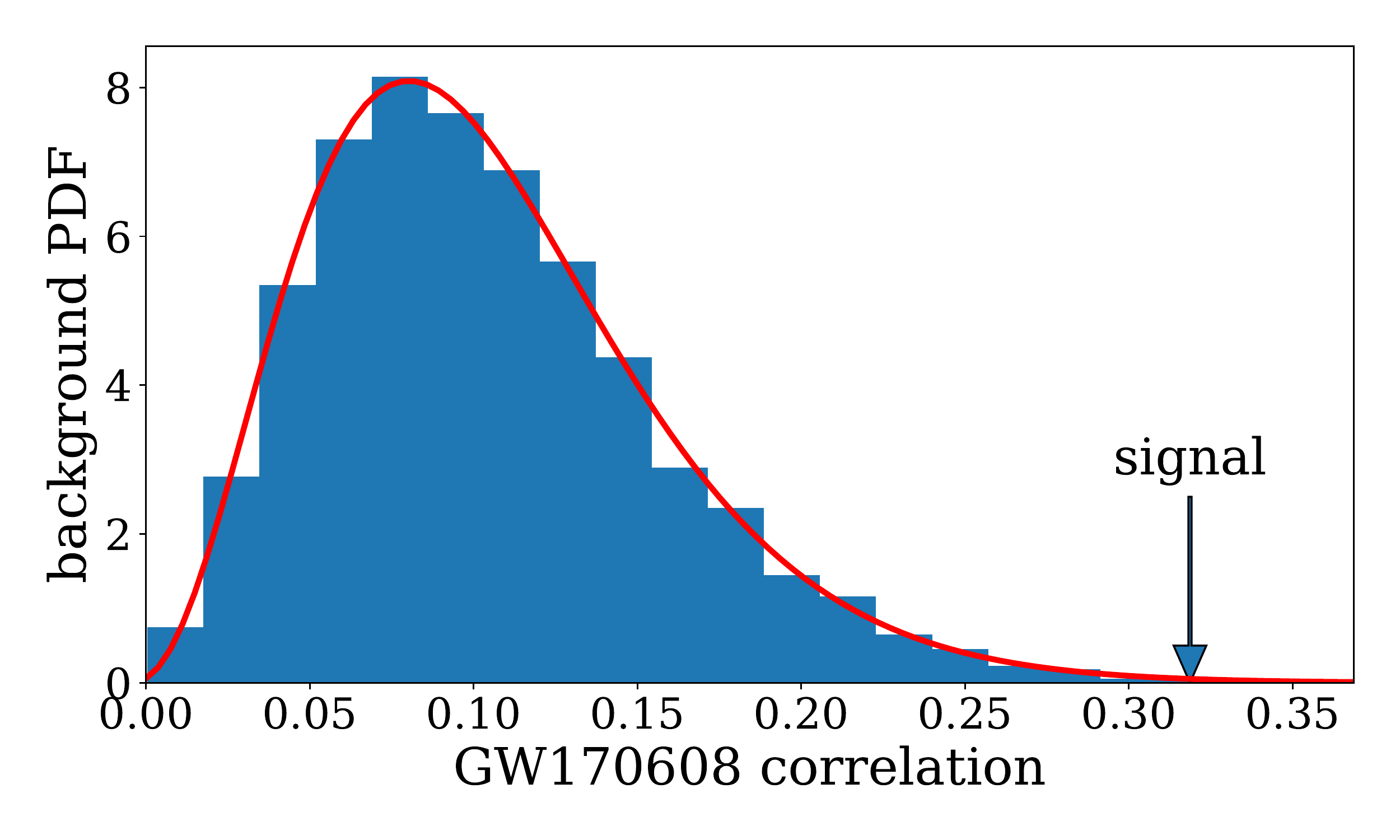}
\end{minipage}
\end{table}
\begin{table}[h!]
\raisebox{15ex}{{FIG.~9.~GW151012}}\hspace{5ex}\begin{minipage}[b]{0.58\linewidth}
\centering
\includegraphics[width=\textwidth]{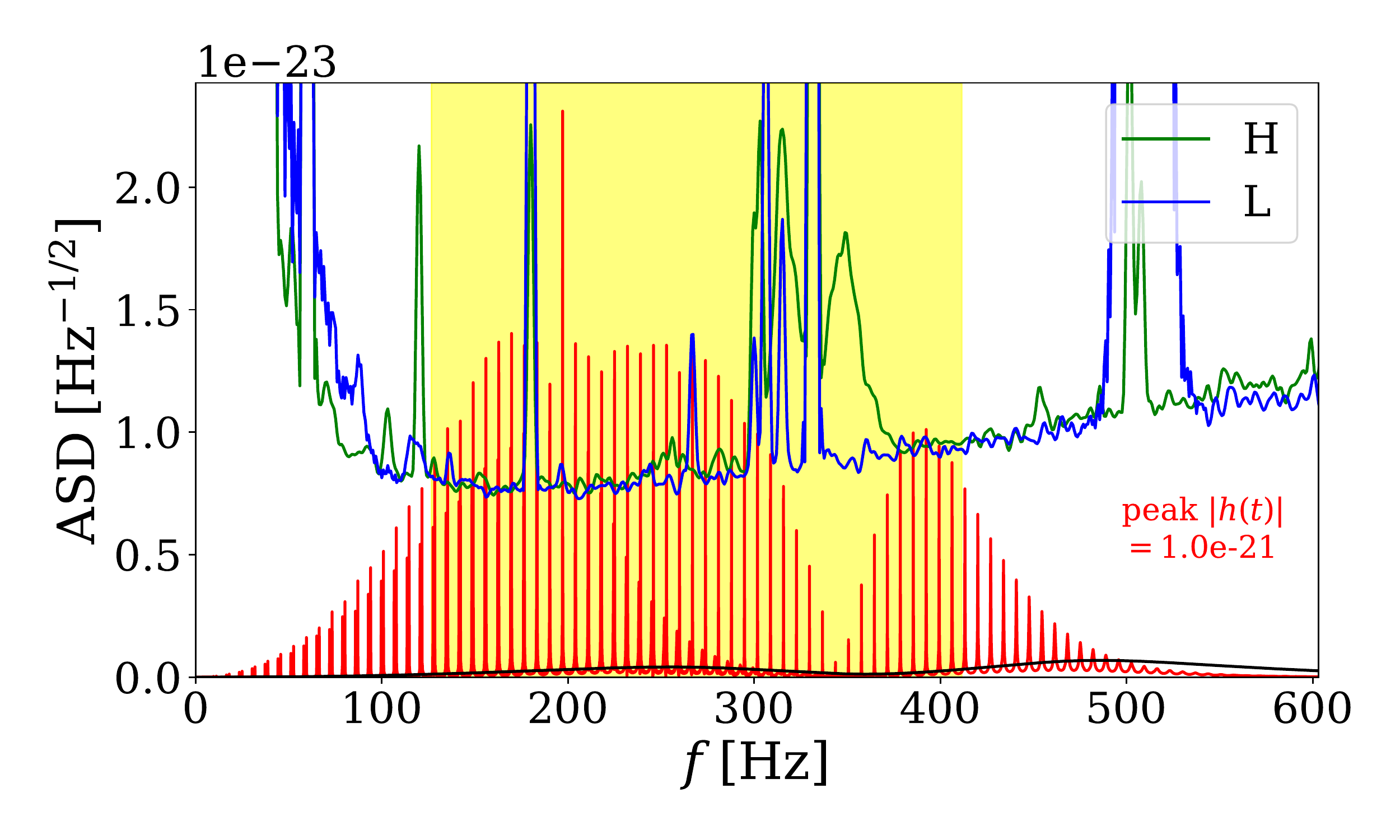}
\end{minipage}\\
\raisebox{15ex}{\begin{minipage}[b]{0.2\linewidth}\centering
\renewcommand{\arraystretch}{1.5}\small
$\begin{array}{|c|c|}
\hline
N_E&160\\
\hline
\Delta f&6.8802{\rm Hz}\\
\hline
\displaystyle\frac{\Delta t}{M}&826\\[5pt]
\hline
\chi&2/3\\
\hline
\end{array}$
\end{minipage}}
\hspace{0.5cm}
\begin{minipage}[b]{0.52\linewidth}
\centering
\includegraphics[width=\textwidth]{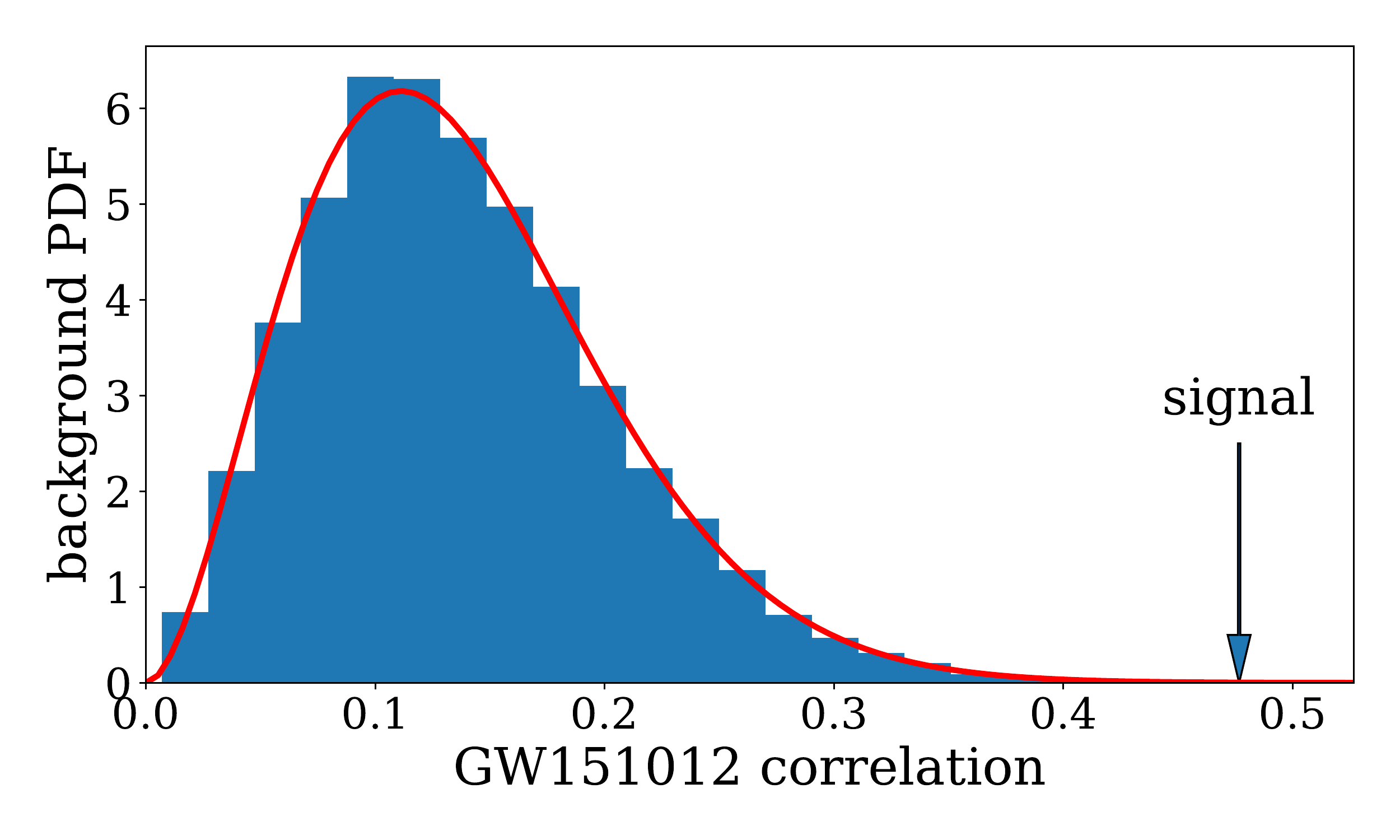}
\end{minipage}
\end{table}

\begin{table}[h!]
\raisebox{15ex}{{FIG.~10.~GW170818}}\hspace{5ex}\begin{minipage}[b]{0.58\linewidth}
\centering
\includegraphics[width=\textwidth]{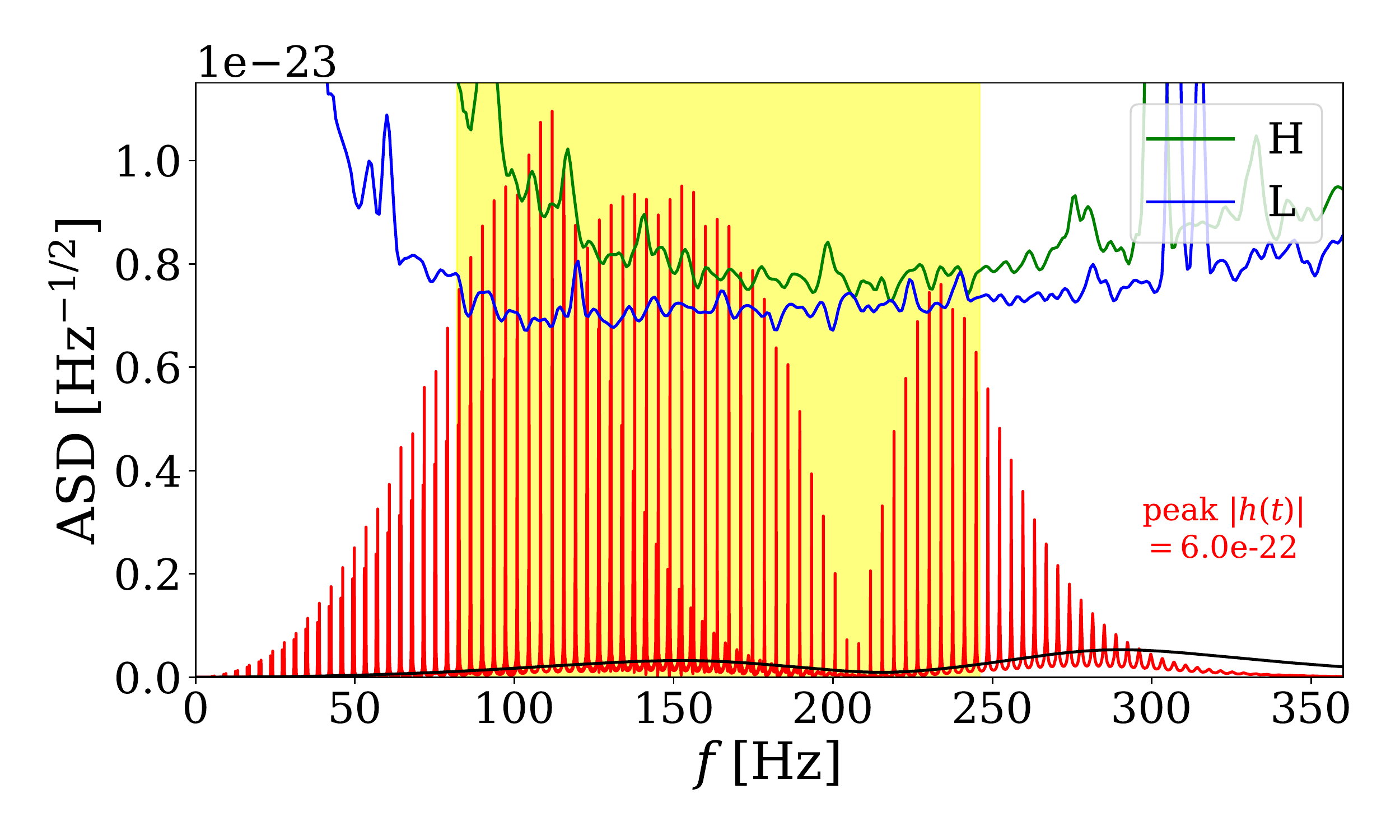}
\end{minipage}\\
\raisebox{15ex}{\begin{minipage}[b]{0.2\linewidth}\centering
\renewcommand{\arraystretch}{1.5}\small
$\begin{array}{|c|c|}
\hline
N_E&140\\
\hline
\Delta f&3.638{\rm Hz}\\
\hline
\displaystyle\frac{\Delta t}{M}&933\\[5pt]
\hline
\chi&2/3\\
\hline
\end{array}$
\end{minipage}}
\hspace{0.5cm}
\begin{minipage}[b]{0.52\linewidth}
\centering
\includegraphics[width=\textwidth]{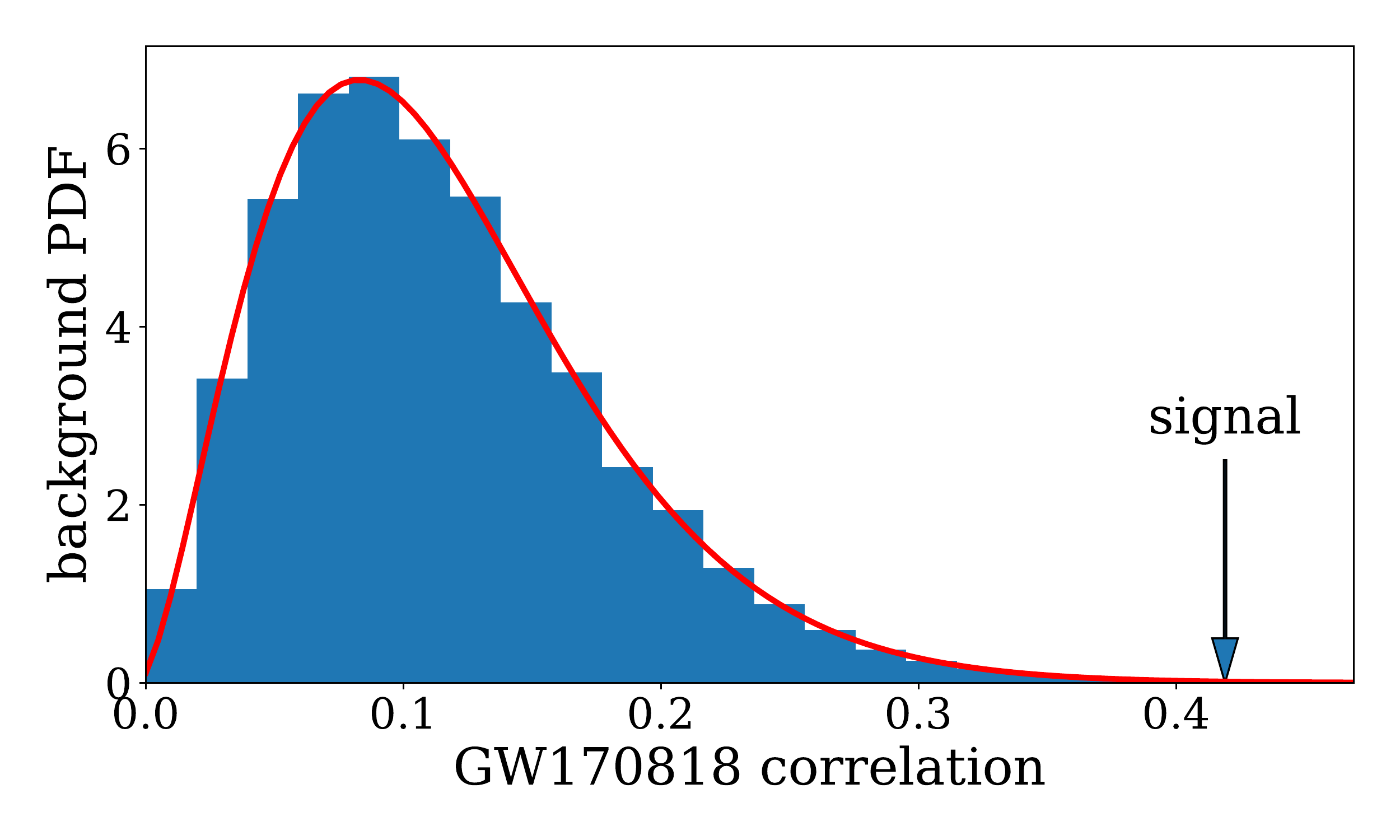}
\end{minipage}
\end{table}
\begin{table}[h!]
\raisebox{15ex}{{FIG.~11.~GW151226}}\hspace{5ex}\begin{minipage}[b]{0.58\linewidth}
\centering
\includegraphics[width=\textwidth]{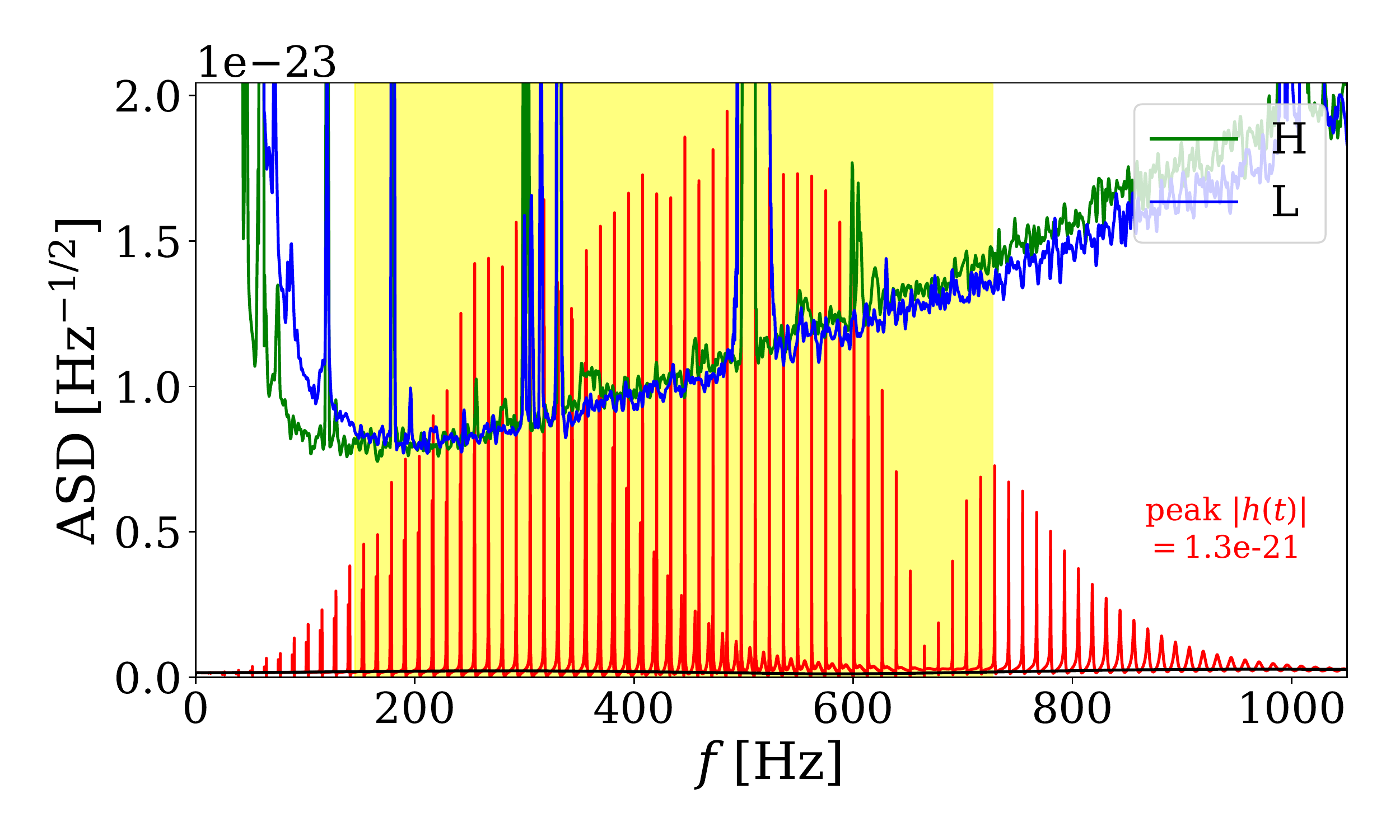}
\end{minipage}\\
\raisebox{15ex}{\begin{minipage}[b]{0.2\linewidth}\centering
\renewcommand{\arraystretch}{1.5}\small
$\begin{array}{|c|c|}
\hline
N_E&270\\
\hline
\Delta f&12.644{\rm Hz}\\
\hline
\displaystyle\frac{\Delta t}{M}&783\\[5pt]
\hline
\chi&0.72\\
\hline
\end{array}$
\end{minipage}}
\hspace{0.5cm}
\begin{minipage}[b]{0.52\linewidth}
\centering
\includegraphics[width=\textwidth]{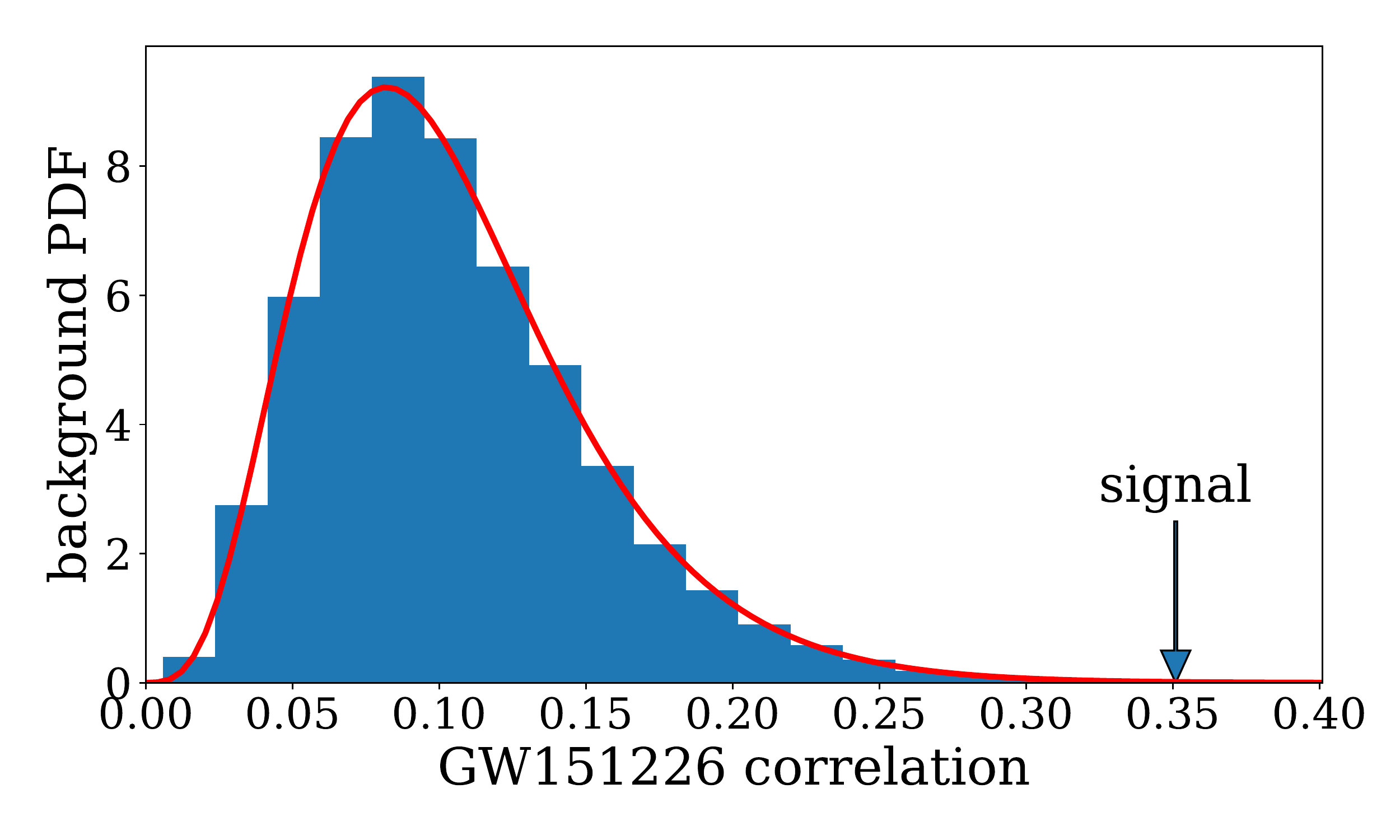}
\end{minipage}
\end{table}

\begin{table}[h!]
\raisebox{15ex}{{FIG.~12.~GW170809}}\hspace{5ex}\begin{minipage}[b]{0.58\linewidth}
\centering
\includegraphics[width=\textwidth]{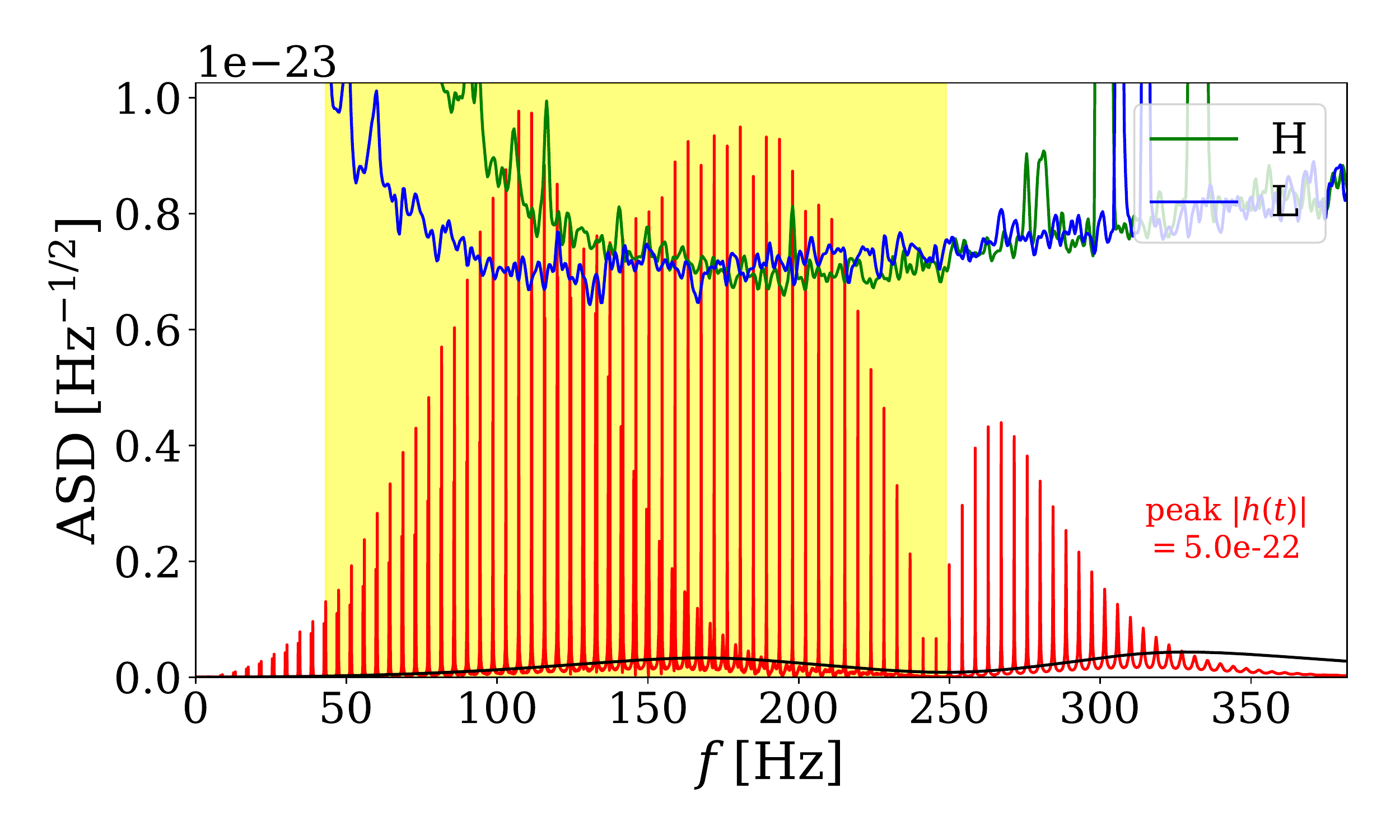}
\end{minipage}\\
\raisebox{15ex}{\begin{minipage}[b]{0.2\linewidth}\centering
\renewcommand{\arraystretch}{1.5}\small
$\begin{array}{|c|c|}
\hline
N_E&170\\
\hline
\Delta f&4.26{\rm Hz}\\
\hline
\displaystyle\frac{\Delta t}{M}&845\\[5pt]
\hline
\chi&0.72\\
\hline
\end{array}$
\end{minipage}}
\hspace{0.5cm}
\begin{minipage}[b]{0.52\linewidth}
\centering
\includegraphics[width=\textwidth]{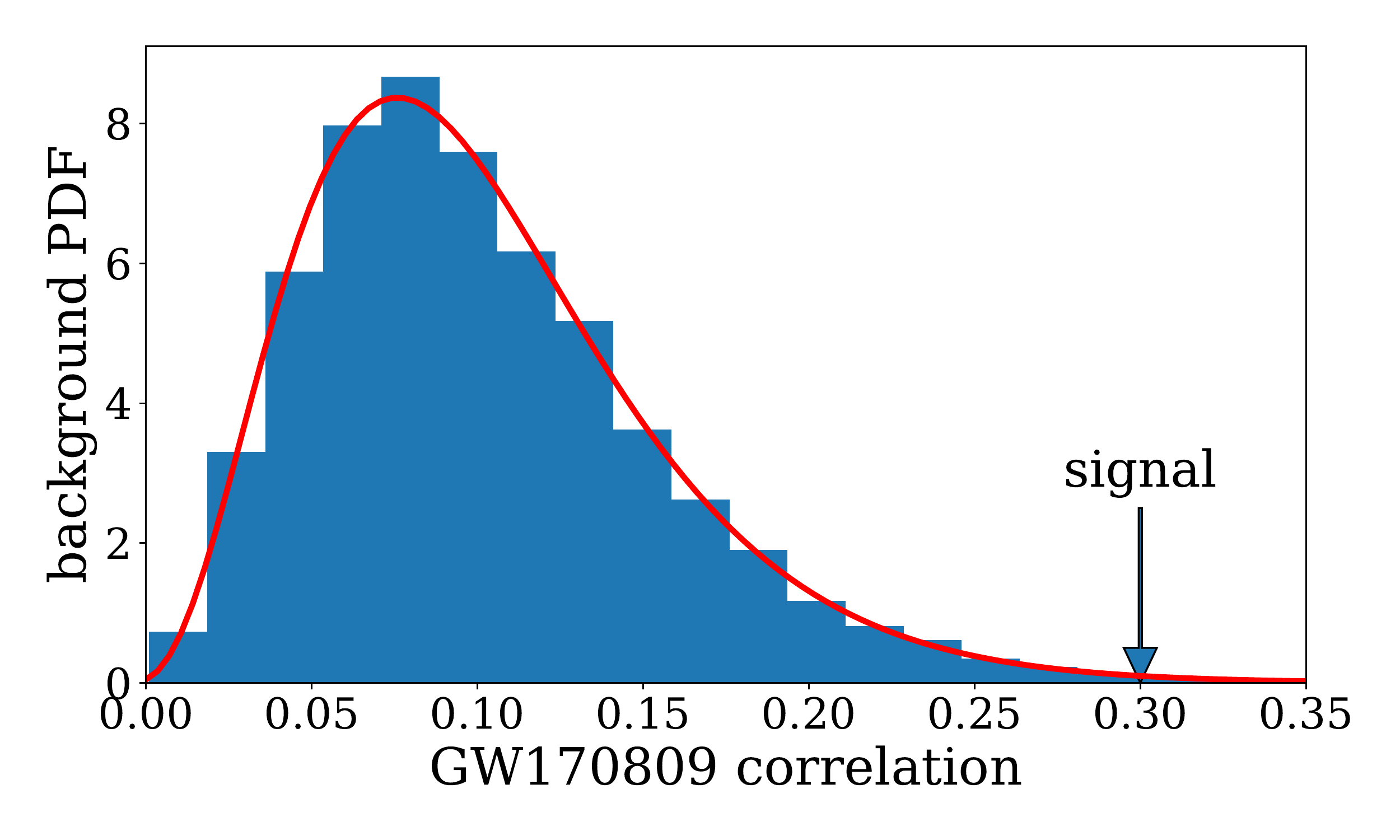}
\end{minipage}
\end{table}
\begin{table}[h!]
\raisebox{15ex}{{FIG.~13.~GW170814}}\hspace{5ex}\begin{minipage}[b]{0.58\linewidth}
\centering
\includegraphics[width=\textwidth]{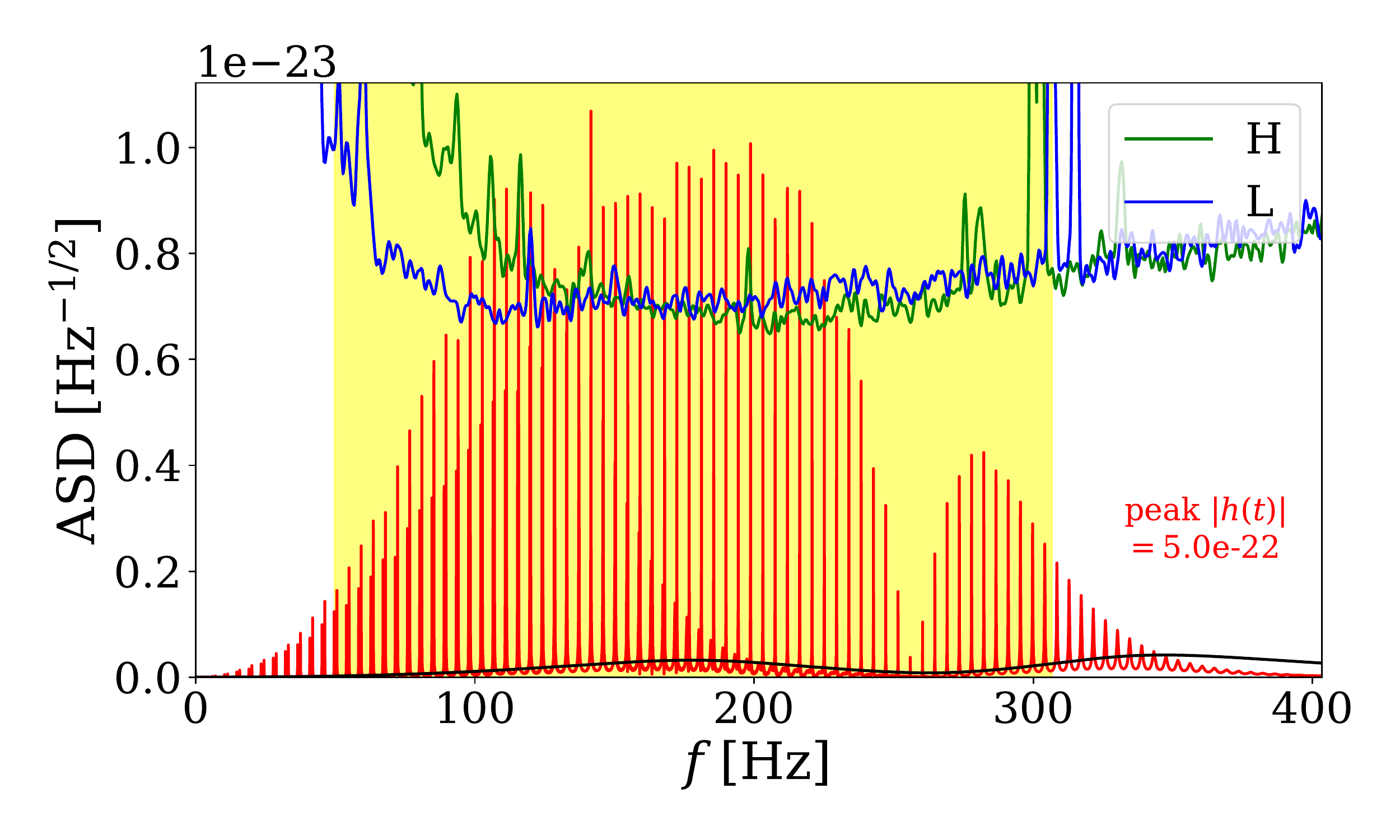}
\end{minipage}\\
\raisebox{15ex}{\begin{minipage}[b]{0.2\linewidth}\centering
\renewcommand{\arraystretch}{1.5}\small
$\begin{array}{|c|c|}
\hline
N_E&200\\
\hline
\Delta f&4.329{\rm Hz}\\
\hline
\displaystyle\frac{\Delta t}{M}&878\\[5pt]
\hline
\chi&0.72\\
\hline
\end{array}$
\end{minipage}}
\hspace{0.5cm}
\begin{minipage}[b]{0.52\linewidth}
\centering
\includegraphics[width=\textwidth]{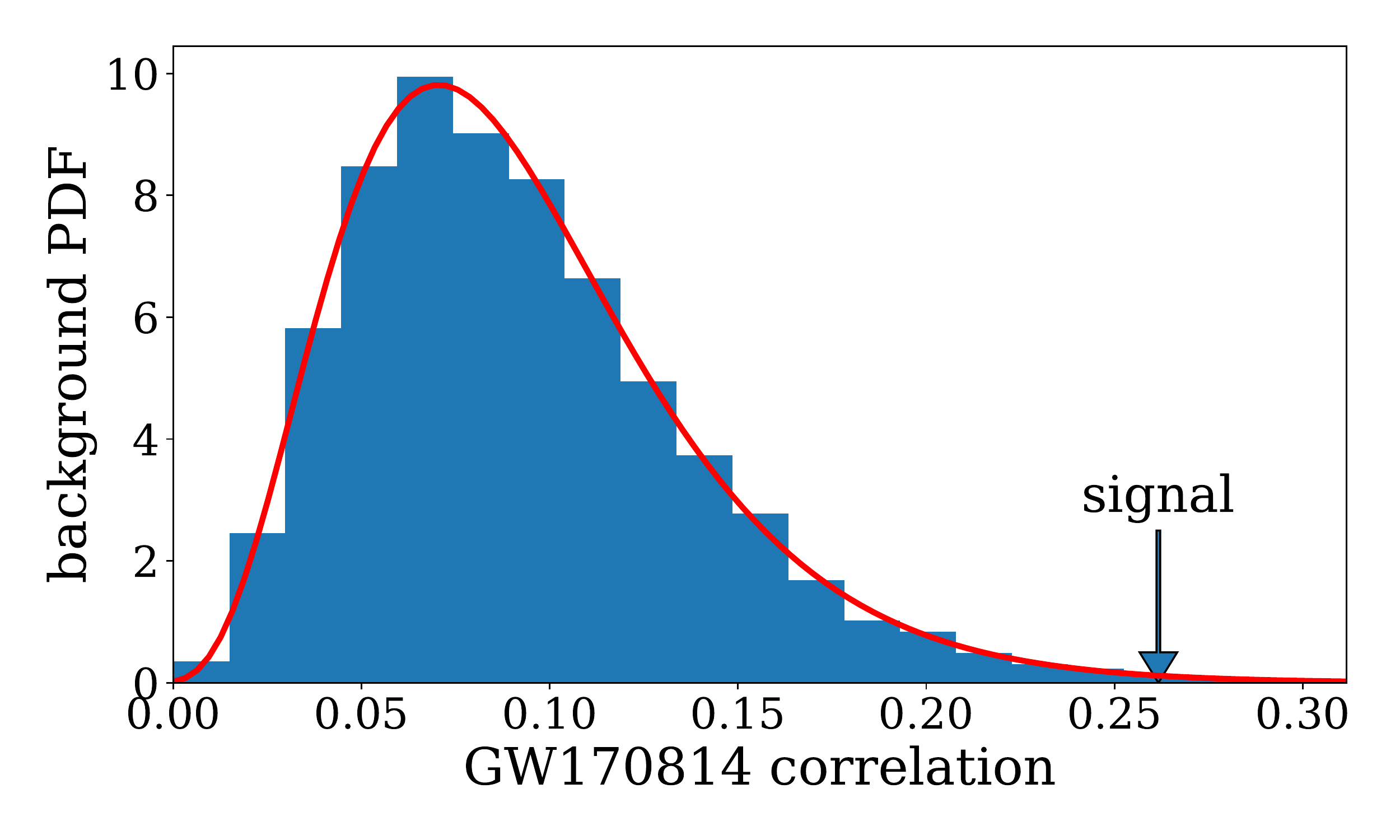}
\end{minipage}
\end{table}

\begin{table}[h!]
\raisebox{15ex}{{FIG.~14.~GW170823}}\hspace{5ex}\begin{minipage}[b]{0.58\linewidth}
\centering
\includegraphics[width=\textwidth]{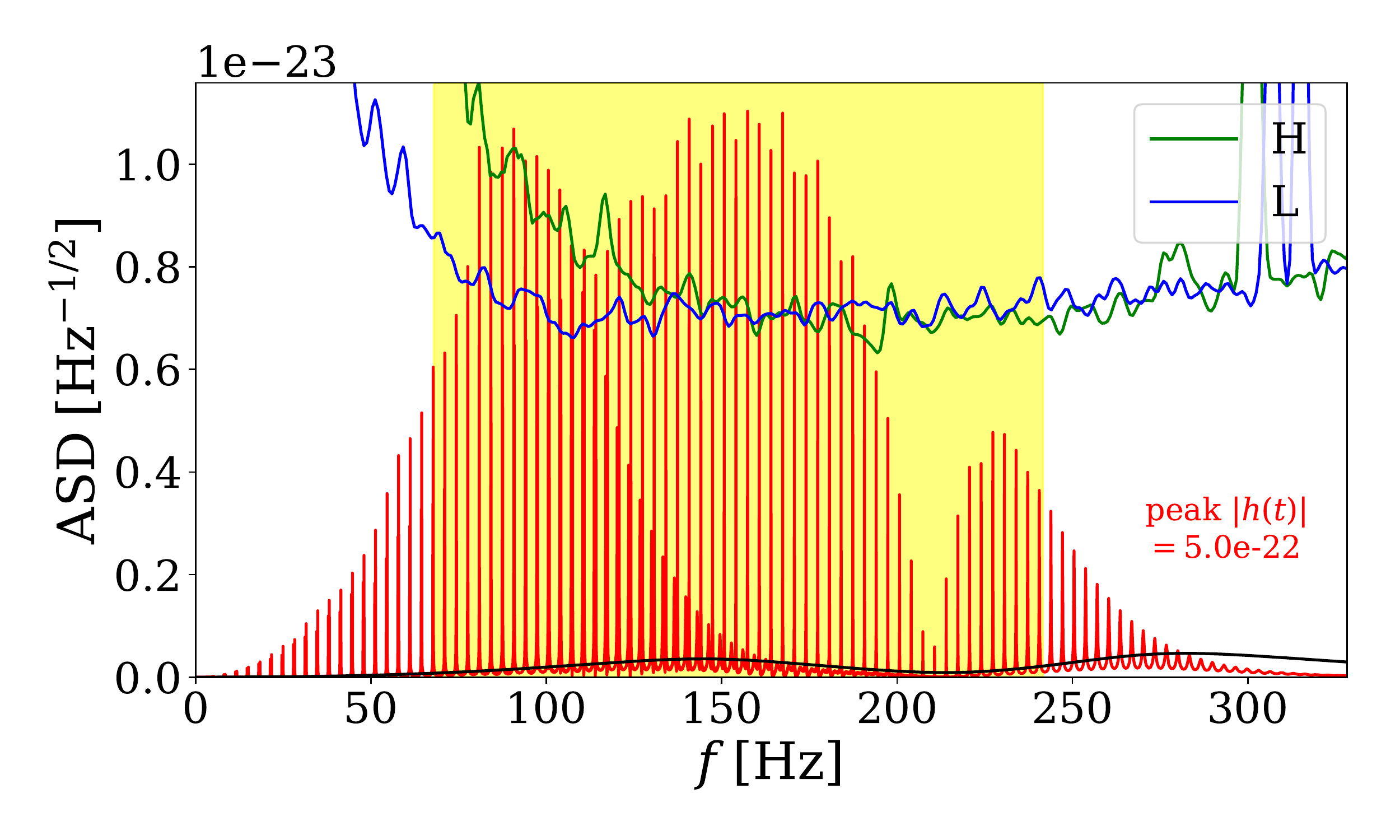}
\end{minipage}\\
\raisebox{15ex}{\begin{minipage}[b]{0.2\linewidth}\centering
\renewcommand{\arraystretch}{1.5}\small
$\begin{array}{|c|c|}
\hline
N_E&200\\
\hline
\Delta f&3.283{\rm Hz}\\
\hline
\displaystyle\frac{\Delta t}{M}&942\\[5pt]
\hline
\chi&0.72\\
\hline
\end{array}$
\end{minipage}}
\hspace{0.5cm}
\begin{minipage}[b]{0.52\linewidth}
\centering
\includegraphics[width=\textwidth]{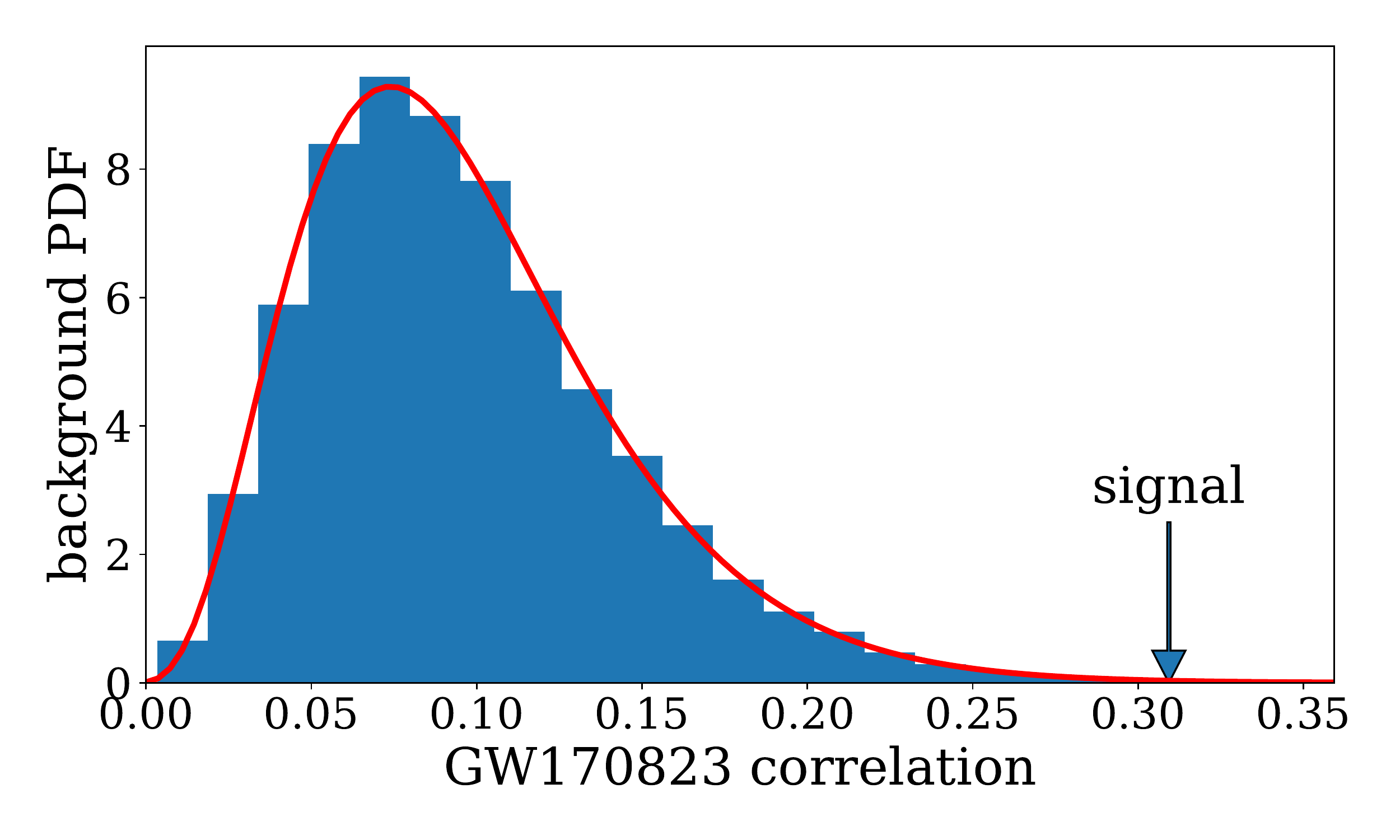}
\end{minipage}
\end{table}

\begin{table}[h!]
\hspace{0.2cm}
\raisebox{14ex}{\begin{minipage}[b]{0.3\linewidth}\centering
{FIG.~15.~GW170729}\\
\renewcommand{\arraystretch}{1.5}\small
\vspace{2ex}$\begin{array}{|c|c|}
\hline
N_E&180/170\\
\hline
\Delta f&2.0441/2.0443{\rm Hz}\\
\hline
\displaystyle\frac{\Delta t}{M}&1240\\[5pt]
\hline
\chi&0.81\\
\hline
\end{array}$
\end{minipage}}
\hspace{0.2cm}\begin{minipage}[b]{0.58\linewidth}
\centering
\includegraphics[width=\textwidth]{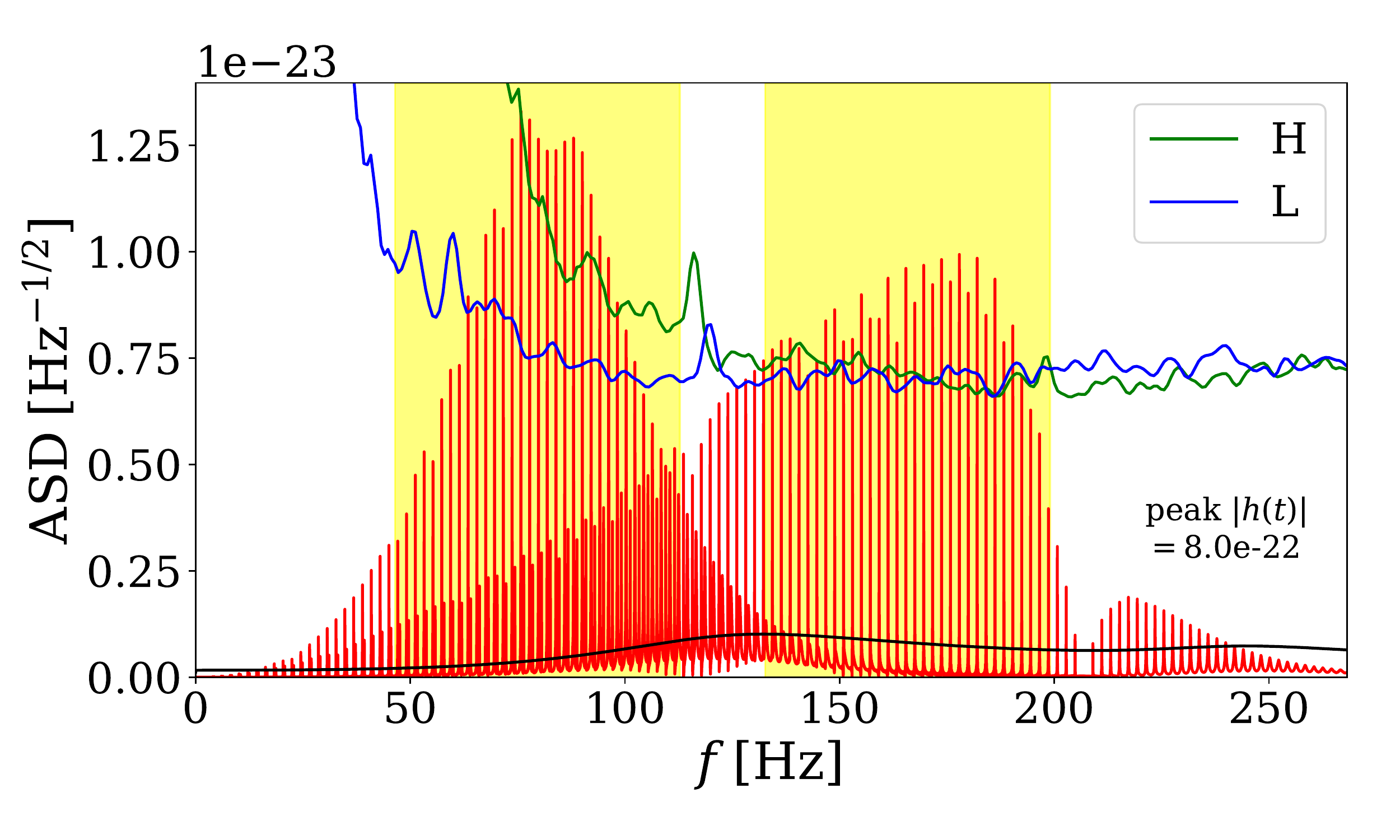}
\end{minipage}
\begin{minipage}[b]{\linewidth}
\centering
\includegraphics[width=.49\textwidth]{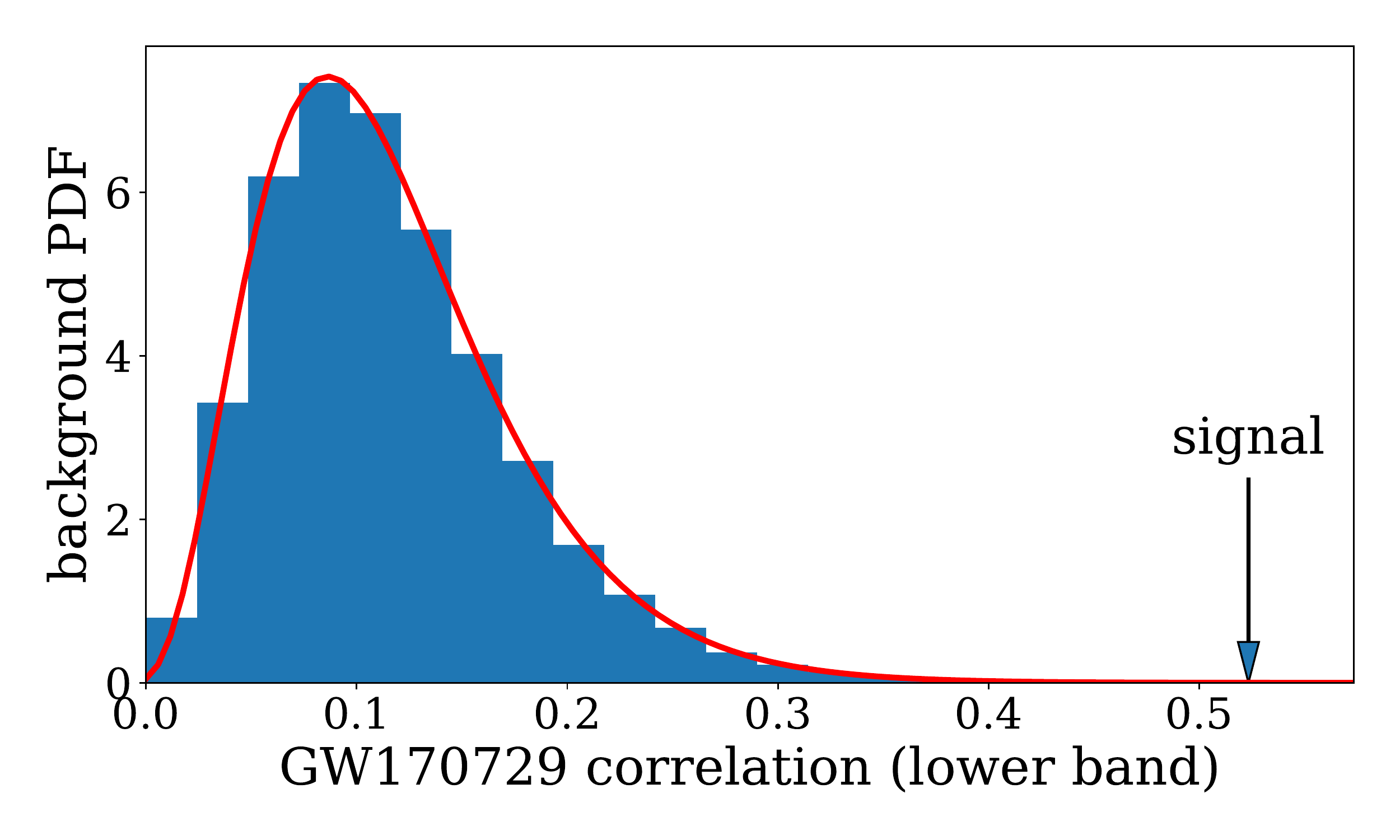}
\includegraphics[width=.49\textwidth]{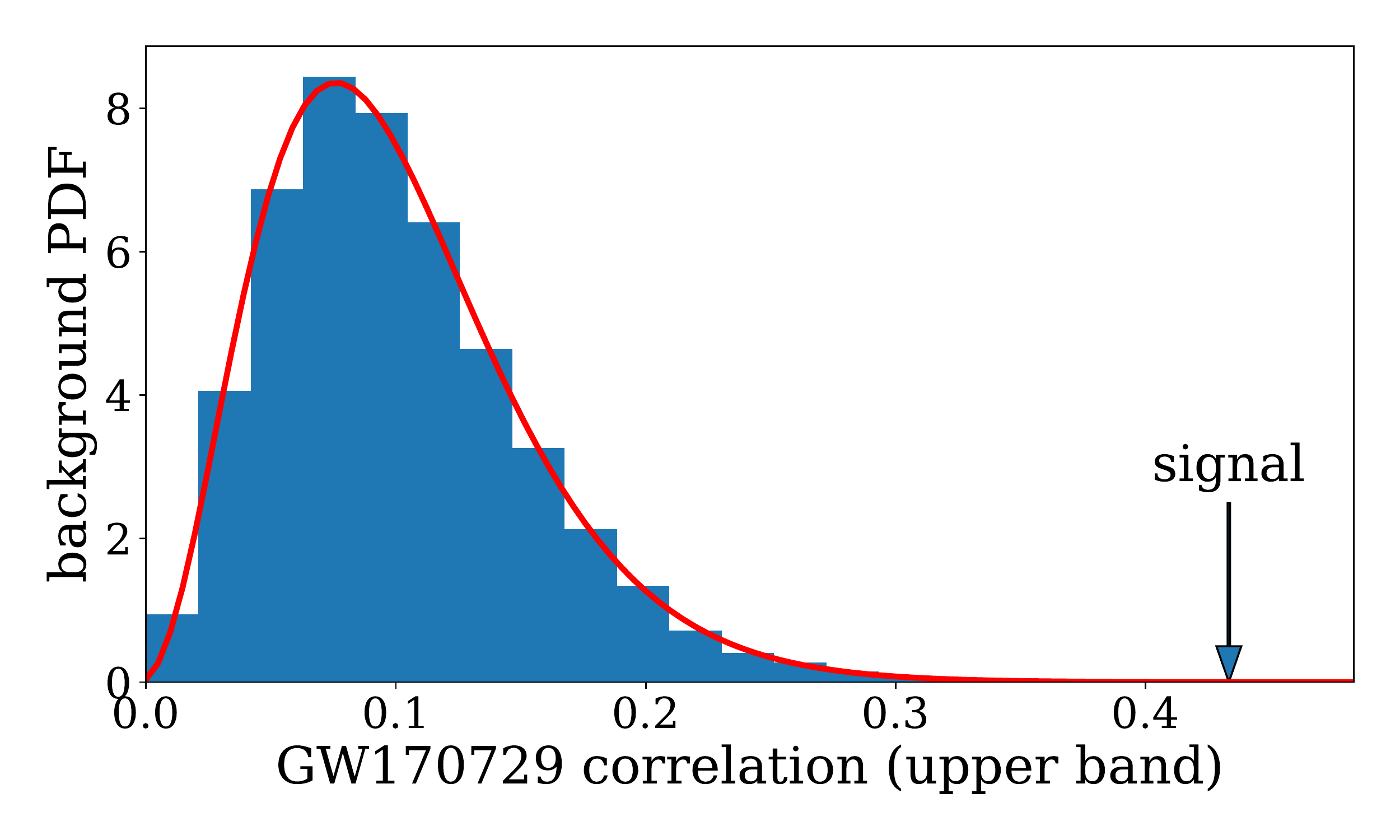}
\end{minipage}
\end{table}

\begin{table}[h!]
\raisebox{15ex}{{FIG.~16.~GW170817}}\hspace{5ex}\begin{minipage}[b]{0.58\linewidth}
\centering
\includegraphics[width=\textwidth]{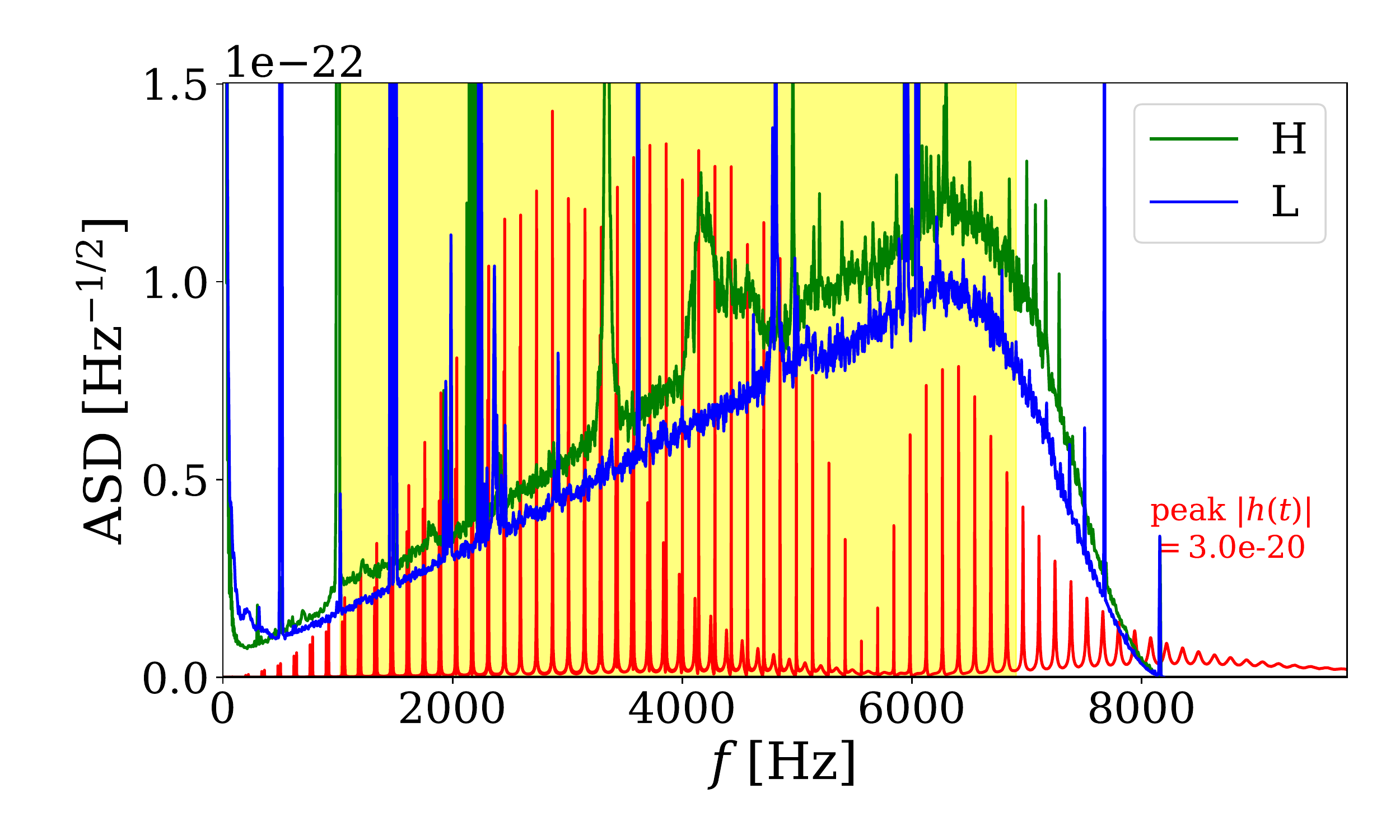}
\end{minipage}\\
\raisebox{14ex}{\begin{minipage}[b]{0.3\linewidth}\centering
\renewcommand{\arraystretch}{1.5}\small
$\begin{array}{|c|c|}
\hline
N_E&250\\
\hline
\Delta f&139.06{\rm Hz}\\
\hline
M\textrm{ (chosen)}&2.2M_\odot\\
\hline
\chi\textrm{ (chosen)}&2/3\\
\hline
\end{array}$
\end{minipage}}
\hspace{0.5cm}
\begin{minipage}[b]{0.52\linewidth}
\centering
\includegraphics[width=\textwidth]{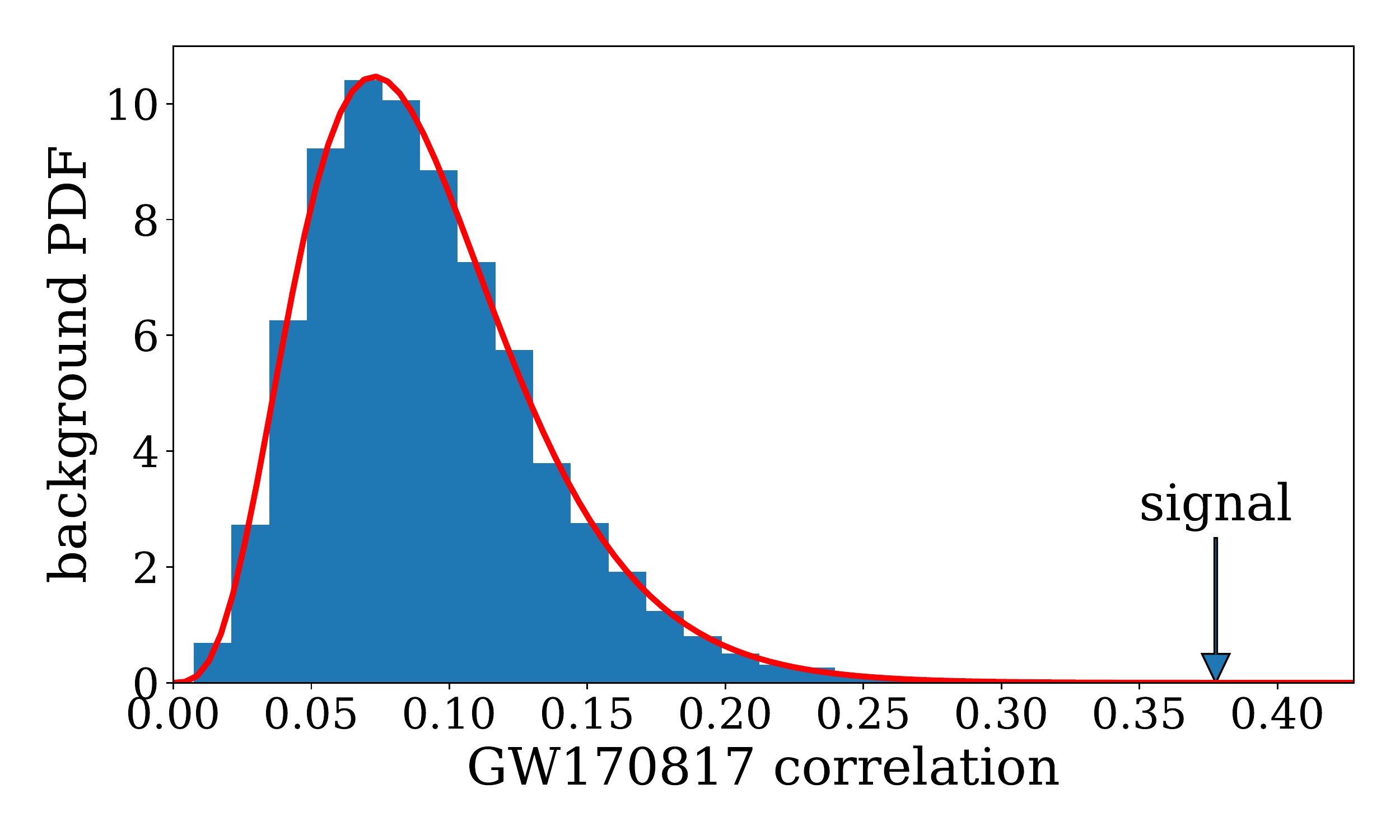}
\end{minipage}
\end{table}

\begin{acknowledgements}
We thank the organizers of Cosmological Frontiers in Fundamental Physics 2019, at the Perimeter Institute September 3-6, where a version of this work was presented. This research was supported in part by the Natural Sciences and Engineering Research Council of Canada.

This research has made use of data, software and/or web tools obtained from the Gravitational Wave Open Science Center (https://www.gw-openscience.org), a service of LIGO Laboratory, the LIGO Scientific Collaboration and the Virgo Collaboration. LIGO is funded by the U.S. National Science Foundation. Virgo is funded by the French Centre National de Recherche Scientifique (CNRS), the Italian Istituto Nazionale della Fisica Nucleare (INFN) and the Dutch Nikhef, with contributions by Polish and Hungarian institutes.
\end{acknowledgements}

\end{document}